\documentclass[useAMS,usenatbib,onecolumn]{mn2e}
\usepackage{enumitem}
\usepackage{graphicx}
\usepackage{caption}
\usepackage{amssymb}
\usepackage{amsmath}
\usepackage{multirow}
\usepackage{mathrsfs}
\usepackage{bbding}
\usepackage{float}
\usepackage{color}
\usepackage[latin9]{inputenc}

\newcommand{\beq}{\begin{equation}}
\newcommand{\eeq}{\end{equation}}
\newcommand{\ignore}[1]{}
\newcommand{\be}{\begin{equation}} \newcommand{\ee}{\end{equation}}
\newcommand{\bea}{\begin{eqnarray}} \newcommand{\eea}{\end{eqnarray}}

\title[Dipole  Anisotropy in Integrated Linearly Polarized Flux Density in NVSS Data]
      {Dipole  Anisotropy in Integrated Linearly Polarized Flux Density in NVSS Data}
\author[Prabhakar Tiwari and Pankaj Jain]
       {Prabhakar Tiwari and Pankaj Jain\\
        Department of Physics, Indian Institute of Technology, Kanpur - 208016, India}

\begin{document}
\maketitle
\begin{abstract}
We study the dipole anisotropy in integrated linearly polarized flux density 
in NRAO VLA Sky Survey (NVSS). We extract the anisotropy parameters 
in the number counts, number counts weighted by polarization observables,
i.e. polarized flux ($P$) 
and degree of polarization ($p$). 
We consider data with several different cuts
on the flux density, $S>10,20,30,40,50,75$ mJy. 
For studies with polarized flux we impose the additional cut
 $0.5<P<100$ mJy. Similarly for degree of polarization 
we impose the cut, $0.01<p<1$. 
We find a very significant signal of dipole,
both in number counts and $P$ or $p$ weighted number counts. 
The polar angle, $\theta$, of the extracted 
dipole axis, for the case of number counts, shows a significant
dependence on the flux density cut. This dependence indicates the
presence of significant bias in the data. We find that the dipole
parameters for the case of number counts weighted by polarized flux
density are relatively stable. We argue that this parameter is relatively
free of bias and study it in greater detail.   
The observed anisotropy is found to be
much larger in comparison to the CMBR expectations. 
We find that polarization observables show
a much higher level of anisotropy in comparison to pure number counts or
sky brightness.
\end{abstract}
\section{Introduction}

There currently exists considerable evidence in favor of a large scale anisotropy in the Universe with
the preferred axis pointing roughly in the direction of Virgo, close to the CMBR dipole. This includes,
radio \citep{Jain:1998r} and optical polarizations \citep{Hutsemekers:1998,Hutsemekers:2000fv,Jain:2003sg}, 
CMBR quadrupole and octopole \citep{Tegmark:2004} as well as the radio source distribution
and brightness \citep{Blake:2002,Singal:2011,Gibelyou:2012,Rubart:2013,Kothari:2013}. 
The physical reason for these observations is not clear and points towards
a violation of the cosmological principle. It might be possible to generate such an anisotropy
 even within the framework of the inflationary big bang
cosmology. This is due to the modes which get produced during the anisotropic
pre-inflationary phase of cosmic evolution \citep{Aluri:2012}.

A dipole anisotropy is expected in the radio source distribution as well as
sky brightness due to Doppler and aberration effects  which arise
due to our local motion \citep{Ellis:1984}.  
CMBR observations indicate that our velocity relative to the cosmic
rest frame is 369$\pm$ 0.9 km/s  in
the direction, $l=263.99^o\pm 0.14^o$, $b=48.26\pm 0.03^o$ in galactic
coordinates \citep{Kogut:1993,Hinshaw:2009}. The direction parameters in
 J2000 equatorial system are
$RA=167.9^o$, $DEC=-6.93^o$. 
There have been many studies which attempt to extract the resulting
dipole anisotropy in radio data \citep{Baleisis:1998,Blake:2002,Crawford:2009,Singal:2011,Gibelyou:2012,Rubart:2013,Kothari:2013}.
Surprisingly, the  
observed anisotropy in radio data is
found to be much larger \citep{Singal:2011,Gibelyou:2012,Rubart:2013,Kothari:2013} 
in comparison to prediction based on  
CMBR observations. However, the direction of the dipole agrees, within errors,
with the CMBR direction.  
Observations also suggest that the amplitude of 
anisotropy in polarization
observables \citep{Jain:1998r} may be larger in comparison to number counts of
radio sources or sky brightness \citep{Singal:2011,Kothari:2013}.  
We point out that diffuse x-ray background also indicates a dipole 
anisotropy \citep{Boughn:2002}, whose direction and amplitude are found to
be consistent with the CMBR dipole.

In the present paper we study the dipole anisotropy in radio polarization data 
 using the NRAO VLA Sky Survey (NVSS) \citep{Condon:1998}. 
We study the anisotropy in number counts after imposing a
cut on the polarization flux ($P$) such that $0.5$ mJy $<P<$ 100 mJy and
independently with a cut on degree of polarization ($p$), such that
$0.01<p<1$. We also study the anisotropy in number counts weighted by
these polarization observables.
The polarization cuts are imposed to select only 
significantly polarized sources. 
An important concern in the present study is the possible presence
of bias in number counts as a function of declination or the polar angle. 
Such a bias is known to exist in the case of number counts of full
data, which includes both polarized and unpolarized sources \citep{Blake:2002}.
This bias gets much reduced if we impose a cut on the flux density, $S$,
of sources, such that, $S<15$
mJy. This is indicated by the fact that the extracted dipole vector
does not show a strong dependence on the lower limit on the flux density,
as long as this limit is greater than 15 mJy \citep{Kothari:2013}. 
This declination dependent bias might be even larger in the 
case of polarized sources. 
We pay careful attention to this problem and identify an observable
which is relatively stable to different cuts imposed on data.
Our results suggest that 
the most reliable variable is the number counts weighted by polarization
flux density. Hence we study it in considerable detail after imposing   
more stringent cuts on polarized flux and error in 
polarized flux.

We follow the procedure used by \cite{Kothari:2013}
and extract the dipole by expanding the observables in spherical 
harmonics. The significance of dipole is calculated by comparing 
the dipole power of data with 
that of 10000 randomly generated isotropic samples. 
We consider several cuts $S>10,20,30,40,50, 75$ mJy on flux density, $S$. 
Due to aberration and Doppler effects we expect that 
the number counts as well as the number counts weighted by $P$ or $p$ 
should display a dipole anisotropy.
We determine the expected kinematic dipole and extract the local speed in 
these cases. 

\begin{figure}
\includegraphics[width=3.5in,angle=0]{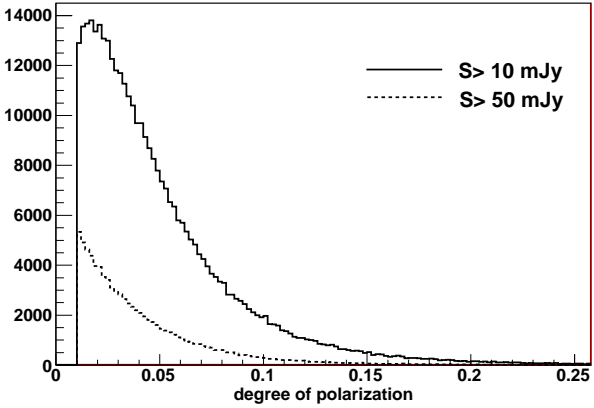}
\caption{The distribution of degree of polarization for NVSS data after 
imposing cuts on the flux density, $S>10$ mJy (solid line) and
$S>50$ mJy (dotted line). Here we have also imposed the cut, $p>0.01$.}
\label{fig:dop}
\end{figure}

\section{The Data}
\label{sc:data}
We use data from the NRAO VLA Sky Survey (NVSS) \citep{Condon:1998}. It is a radio continuum 
survey at 1.4 GHz, covering the sky north of $\delta=-40^{o}$,
where $\delta$ is the J2000 declination. The data product 
contains a catalogue of 1773484 radio sources. We follow \cite{Blake:2002} to impose 
different cuts on intensity. 
We remove Galactic contamination by masking the Galactic plane within latitude $|b|<15^\circ$.
Furthermore, we remove the clustering dipole \citep{Blake:2002} by 
removing the sources within 30 arcsec of known nearby galaxies as listed in 
\cite{Saunders:2000} and in third reference catalogue of bright Galaxies  
(RC3) \citep{Vaucouleurs:1991,Corwin:1994}. 
In addition we also impose cuts on the polarization observables. 
The histogram of the degree of polarization is given
in Fig. \ref{fig:dop}.   

\begin{figure}
\includegraphics[width=3.5in,angle=0]{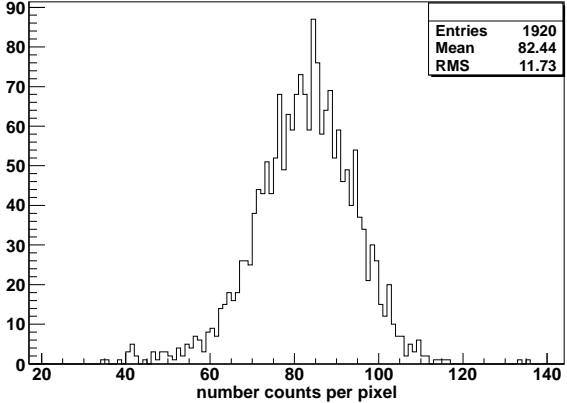}
\caption{The distribution of number counts of significantly polarized 
sources for Nside$=$16.  
Here we have imposed the cuts, $S>20$ mJy and $0.5<P<100$ mJy where $S$ 
is the flux density and $P$ the polarized flux density.
}
\label{fig:pix}
\end{figure}

We use HEALPix\footnote{http://healpix.jpl.nasa.gov/} equal area pixelization  scheme (Nside$=$16) to 
bin the sources in equal area pixels of $\sim 3.7^{\circ}$. The distribution
 of the number of significantly polarized sources per 
pixel in shown in Fig. \ref{fig:pix}. In this figure we have imposed the cuts, 
$S>20$ mJy and
$0.5<P<100$ mJy. 

\section{Procedure}
\label{sc:analyses}
The polarized flux density,
$P(\theta,\phi)=\sqrt{Q^2 +U^2}$ and degree of
polarization, $p(\theta,\phi)=\sqrt{Q^2 +U^2}/S$, where $Q$ and
$U$ are the standard Stokes parameters. 
Here $(\theta,\phi)$ represent the spherical polar coordinates of the source. 
We use J2000 equatorial coordinate system, such that $RA=\phi$ and $Dec=90-
\theta$. 
Both $P$ and $p$ are
invariant under rotations about the direction of propagation.
We rewrite these fields as $P(\theta, \phi) = P_{0}(1+ \Theta(\theta, \phi))$,
with a similar equation for $p(\theta,\phi)$. Here 
$\Theta(\theta, \phi)$ represents the real space fluctuations of the polarization 
field. To study the correlations of this polarization field we expand 
it in spherical harmonics,
\beq
\label{eq:p2a}
\Theta(\theta, \phi) = \sum_{l=1}^{\infty} \sum_{m=-l}^{+l} a_{lm} Y_{lm}(\theta,\phi),
\eeq
where $Y_{lm}(\theta,\phi)$ are the usual spherical harmonics and $a_{lm}$ the expansion parameters. 
The power for each multipole is given by,
\beq
\label{eq:C2alm}
C_l =\frac{1}{(2l+1)} \sum_{m=-l} ^{l} |a_{lm}|^2. 
\eeq
A significant value 
of $C_l$ indicates anisotropy at a scale $\sim (\pi/l)$ radian.
In particular, $C_{1}$ represents the dipole term and its 
relation with  dipole amplitude 
$D$ is \citep{Gibelyou:2012},
\beq
\label{eq:c2d}
C_1 = \frac{4\pi}{9} D^2.
\eeq

Anisotropy in number counts is expected due aberration and Doppler
effects caused by local motion \citep{Ellis:1984}. Let $\nu_o$ and
$\nu_r$ represent the observed and rest frame frequencies respectively. 
These are related by $\nu_o=\nu_r\delta$ where,
$\delta = 1+v\cos\theta/c$, where $v$ is the local speed and 
$\theta$ the angle between the direction of observation and the local
velocity.
The polarized flux density and the degree of polarization 
\citep{Eichendorf:1979,Mesa:2002,Tucci:2004,Jackson:2010} also show 
a dependence on frequency.
We assume that the polarized flux density, $P$, 
can be modelled, approximately, as $P \propto\nu^{-\alpha_P}$.  
The parameter, $\alpha_P$, is so far unknown for NVSS data. 
Here we assume that $\alpha_P =0.75$, 
same as $\alpha$ for intensity \citep{Blake:2002}. This value may
be revised in future once more data spectral polarization data becomes
available. Small changes in this parameter have minor effect on our
results and do not affect our conclusions.  

Let $S$ represent the total flux density of a source. 
Let $n(\theta,\phi,P,S)$ denote the differential number count per unit solid angle per unit polarized flux density ($P$) per unit total flux density ($S$). 
In the cosmological rest frame, assuming
isotropy, we may model this as a power law in the parameters $P$ and $S$, i.e.,
\begin{equation}
 n_{rest}(\theta,\phi,P_{rest},S_{rest})\equiv {d^3N_{rest}\over d\Omega_{rest} dP_{rest}dS_{rest}} =  {k x x_P} \left({P_{rest}}\right)^{-1-x_P}  
\left({S_{rest} }\right)^{-1-x}
\label{eq:d2N}
\end{equation}
where both $P_{rest}$ and $S_{rest}$ are in units of mJy. 
Here $d^3N_{rest}$ represents the number of sources in a small bin, $d\Omega_{rest} dP_{rest} dS_{rest}$ in the rest frame. Here we have used a 
simple power behaviour assuming that $S$ and $P$ dependence may be
decoupled from one another. These assumptions may not be 
exactly valid.  
The degree of polarization is also known to be anti-correlated with
the flux density \citep{Mesa:2002,Tucci:2004}.   
Furthermore a simple power law may not provide a very good fit to data.
However this model is reasonable to 
get an approximate estimate of the expected kinematic effect.   
More detailed fits, along the lines discussed in \cite{Kothari:2013},
may be explored in future work. 
We obtain the values of $x$ and $x_P$ by fitting the data 
with different cuts. 
Let $d^3N_{obs}$ represent the corresponding
number of sources in the observer frame. We have \citep{Ellis:1984,Kothari:2013}, 
\begin{equation}
 d^3N_{obs} = d^3N_{rest} = n_{rest} d\Omega_{rest}dP_{rest}dS_{rest} = 
kxx_P \left({P_{obs} }\right)^{-1-x_P} \left({S_{obs} }\right)^{-1-x} 
\delta^{2+x(1+\alpha)+x_P(1+\alpha_P)}
dP_{obs} dS_{obs} d\Omega_{obs} 
\end{equation}
where we have used $d\Omega_{obs} = d\Omega_{rest}\delta^{-2}$,
$S\propto \nu^{-\alpha}$, $P\propto \nu^{-\alpha_P}$, $\nu_o=\nu_r\delta$,
$S_{obs} = S_{rest}\delta^{1+\alpha}$,
$P_{obs} = P_{rest}\delta^{1+\alpha_P}$.

The analysis above implies that we expect a dipole 
anisotropy in number counts of significantly polarized sources, given by,
\beq
\label{eq:D_n}
D_{N}(v) = [2+x(1+\alpha)+x_P(1+\alpha_P)](v/c).
\eeq
We point out that here we have assumed that $S$ and $P$ are independent
variables. The terms $x(1+\alpha)$, $x_P(1+\alpha_P)$ arise 
due to the lower limit imposed on $S$ and $P$ respectively. However 
we find that, for the case of real data, 
these two cuts are not independent. In particular a
stringent cut on the flux density also eliminates most of the sources with
low polarization and vice verse. Here we ignore this complication since 
the extracted velocity is not found to be physically relevant. 
As we shall see, the extracted dipole in all likelihood gets a large
contribution from some unknown intrinsic effect.

We next consider the total polarized flux, $P_I$, i.e. the number
counts weighted by the polarized flux density. This can be expressed as,
\beq
\label{eq:P_I}
d^3P_I = Pd^3N
\eeq
In each pixel, this measure is computed by summing over the polarized
flux density of all sources in the pixel. 
Following the procedure described above, we find that the expected 
dipole anisotropy, $D_P$, in this measure due to kinematic effects 
is identical to that expected in pure
number counts, i.e.,
\beq
\label{eq:D_I}
D_{P}(v) = [2+x(1+\alpha)+x_P(1+\alpha_P)](v/c).
\eeq
This is consistent with the result obtained earlier for unpolarized 
observables \citep{Singal:2011,Kothari:2013}, where it was found that
the kinematic effects lead to identical dipole anisotropy in source
counts and sky brightness if we assume a power law form of the differential
number counts, $n(\theta,\phi,S)$.  

We finally generalize this to the case of degree of polarization, $p=P/S$. 
The spectral dependence of $p$ is given by, $p\propto \nu^{-\alpha_P+\alpha}$. 
We also define degree of polarization per unit frequency interval, $p'$. 
This is related to $p$ by, $dp = p'd\nu$. We now repeat the above calculation
using $p'$ as the variable instead of $P$. We find that the dipole in
number counts in this case is given by,
\beq
\label{eq:D_np}
D_{Np}(v) = [2+x(1+\alpha)+x_p(\alpha_p-1)](v/c).
\eeq
where $\alpha_p = \alpha_P-\alpha+1$. We point out that in this case,
$p'_{obs} = p'_{rest}\delta^{\alpha_p-1}$. 
We also consider the dipole anisotropy in number counts weighted by
the degree of polarization. This is defined by Eq.  
\ref{eq:P_I}, with $P$ replaced on the right hand side by $p$, i.e.
$d^3p_I = pd^3N$. 
Our choice, $\alpha_P=\alpha$, implies that the degree of polarization
is independent of frequency. This is in disagreement with the observations,
which indicate an increase in $p$ with frequency \citep{Klein:2003,Tucci:2004}.  
Hence it might be more reasonable to choose a smaller value of $\alpha_P$. 
However, as discussed above, small changes in this parameter have
negligible effect on our conclusions. In any case, a smaller value
of $\alpha_P$ will further enhance the difference between the extracted
local speed and that expected from CMBR observations. 
The amplitude of the dipole anisotropy in the number counts weighted by 
degree of polarization is also given by Eq. \ref{eq:D_np}. 
The corresponding dipole anisotropy is denoted as $D_p$.

\section{Analysis Method and Bias Simulation}
\label{sc:bias}
There are several regions of the sky for which data is not available. In 
particular we do not have data for $\delta<-40^o$ and we  
mask the Galactic plane within 
latitude $|b|<15^\circ$ in order to remove the Galactic contamination. 
We create a full sky map by filling all of these 
empty pixels by randomly generated
isotropic data 
\citep{Kothari:2013}. The random data is generated directly from the
distribution observed for the real data set. For the number counts per 
pixel, this distribution is shown for a particular set of cuts in  
Fig. \ref{fig:pix}. The number of sources in empty pixels are obtained
by randomly allocating the counts in filled pixels to the empty
pixels. This preserves their distribution.  
Once the number count of sources in all the empty pixel is determined,
the polarized flux of these sources is assigned by randomly
extracting the values from real data. 
Hence the data in empty pixels has the same statistical properties as that
in real pixels but does not have the large scale anisotropy, which might
be present in real data. We make a spherical harmonic decomposition of
the full sky map and extract the dipole parameters. The corresponding 
dipole power is denoted by $C_1'$ and the dipole axis parameters $(\theta',\phi')$, where $\theta'$ is the polar angle and $\phi'$ the RA.  
The extracted parameters would depend on the random sample used for filling 
the masked regions. Hence we repeat this process 1000 times. The final 
dipole parameters, $C_1'$, $\theta'$ and $\phi'$ 
are obtained by taking the average over these 1000 realizations. 

The random filling of empty pixels also adds a bias in dipole magnitude and 
direction. Hence the extracted average parameters $C'_1$, $\theta'$ and
$\phi'$ would contain some bias. We determine this bias by simulations. 
We generate full sky 
random samples which lead to 
a dipole anisotropy same as that seen in real data. 
We first generate an isotropic random map by randomly allocating NVSS data to 
different pixels. The procedure followed here is identical to that followed
in filling the empty pixels in real data set, as described above. 
We then add a dipole 
with a magnitude and direction similar to that seen in real data.
Let the input dipole power be denoted by $C_1$, and the axes
parameters, $(\theta,\phi)$.  
We next apply the same mask to the simulated maps as applied to real data. 
The pixels in the masked regions are filled by a procedure
identical to that followed for the real map. 
The difference between the dipole parameters extracted from the simulated map
 and the corresponding input
parameters, gives a measure of the bias generated due to random filling
of the masked sky. This procedure is iterated
till the dipole parameters extracted from the simulated map match those 
of the real map. Hence the dipole parameters extracted from a simulated
map are also equal to $C_1'$, $\theta'$ and $\phi'$, the same as those for
real map. We have verified that the final result is independent of the
initial guess used for the dipole vector. 
We generate 1000 random simulated maps. The output of each such
map is equal to $C_1'$, $\theta'$ and $\phi'$, where as the input 
dipole parameters would differ due to the statistical fluctuations in the
random realizations. The bias factor corresponding to the dipole power 
is defined as, $k^2 = C_1'/C_1$. Similarly
the bias factors for angular variables are defined as, $\Delta\theta=
\theta'-\theta$ and $\Delta\phi = \phi'-\phi$ \citep{Kothari:2013}.

\begin{figure}
\includegraphics[width=3.5in,angle=0]{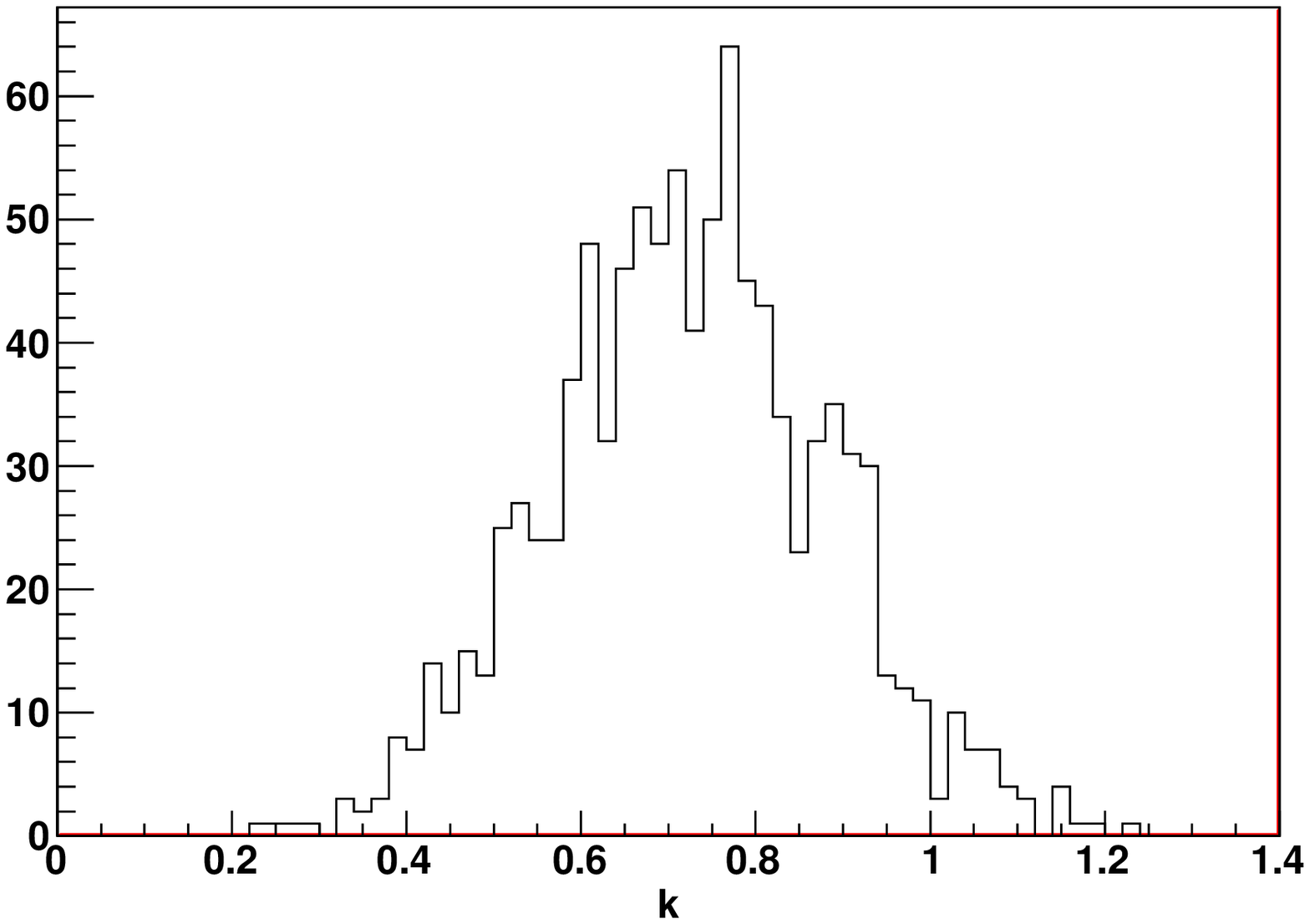}\\
\includegraphics[width=3.5in,angle=0]{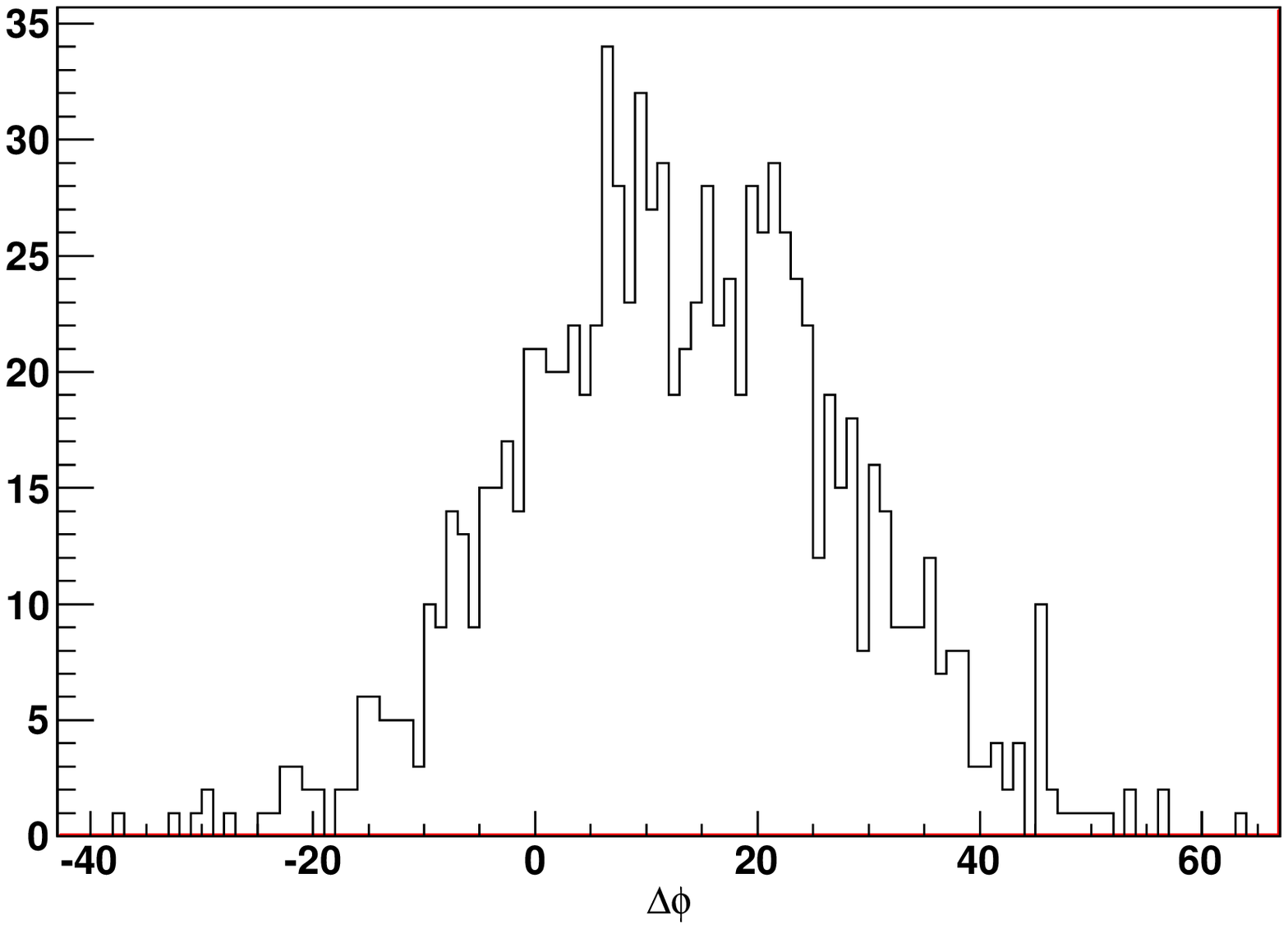}\\
\includegraphics[width=3.5in,angle=0]{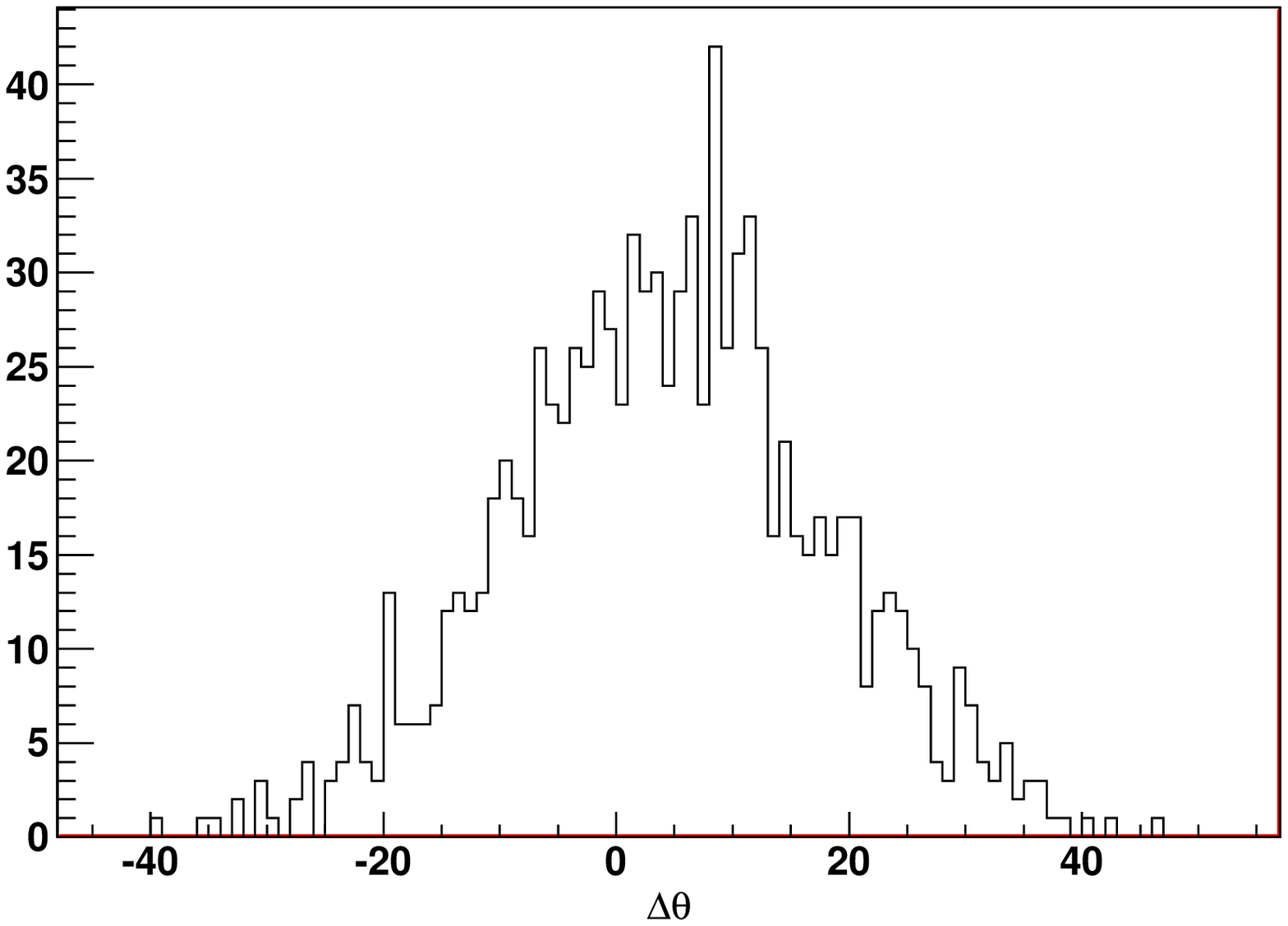}\\
\caption{The distribution of the bias factors 
$k=  \sqrt{C'_1 /C_1}  $, $\Delta \theta(\theta^{'}-\theta)$ 
and $\Delta \phi(\phi^{'}-\phi)$ for number counts of significantly polarized
sources after imposing the cuts
$S_{low} = 20$ mJy and $0.5<P<100$ mJy.} 
\label{fig:c1sim2_PI}
\end{figure}

\begin{figure}
\includegraphics[width=3.5in,angle=0]{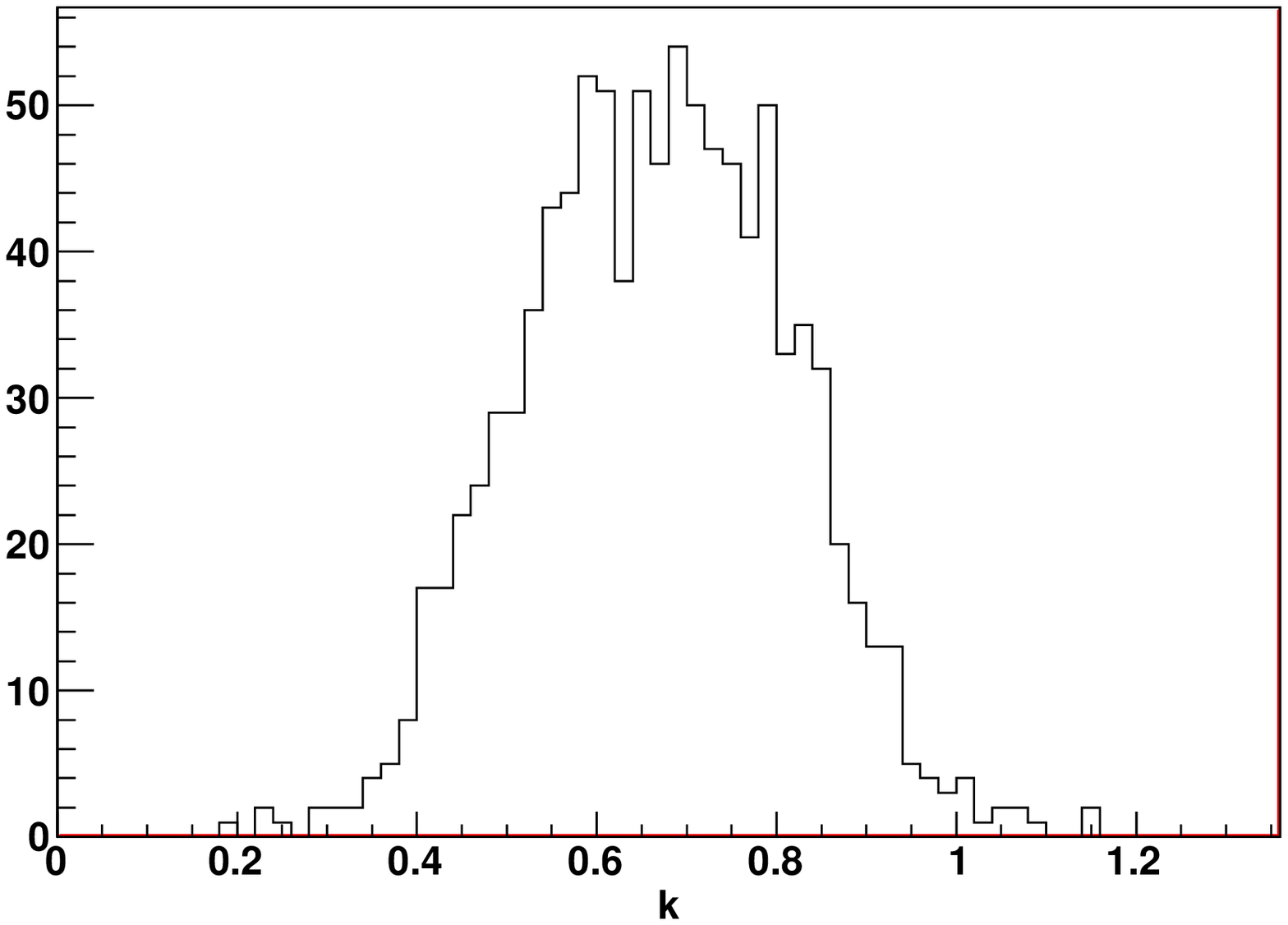}\\
\includegraphics[width=3.5in,angle=0]{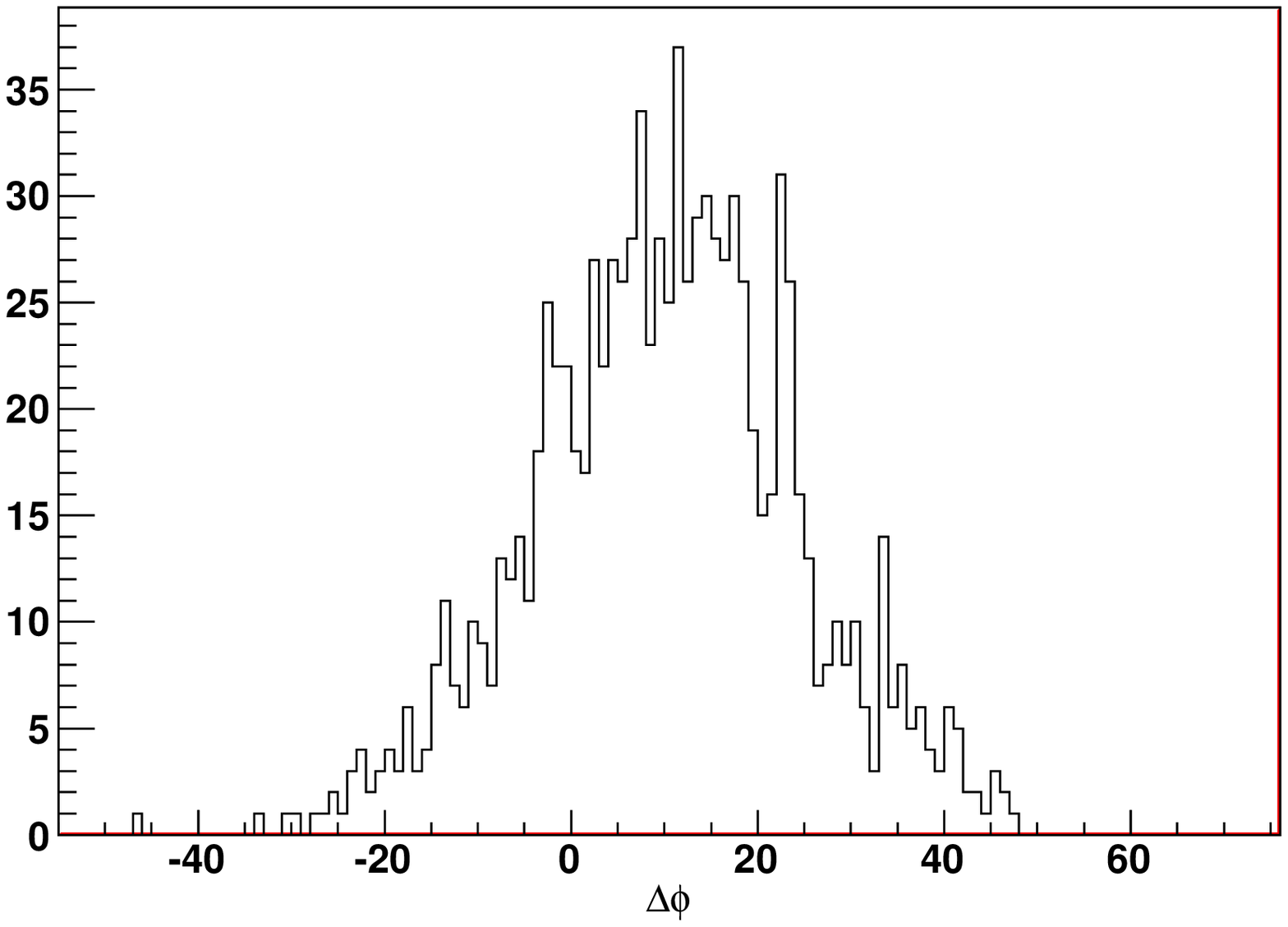} \\
\includegraphics[width=3.5in,angle=0]{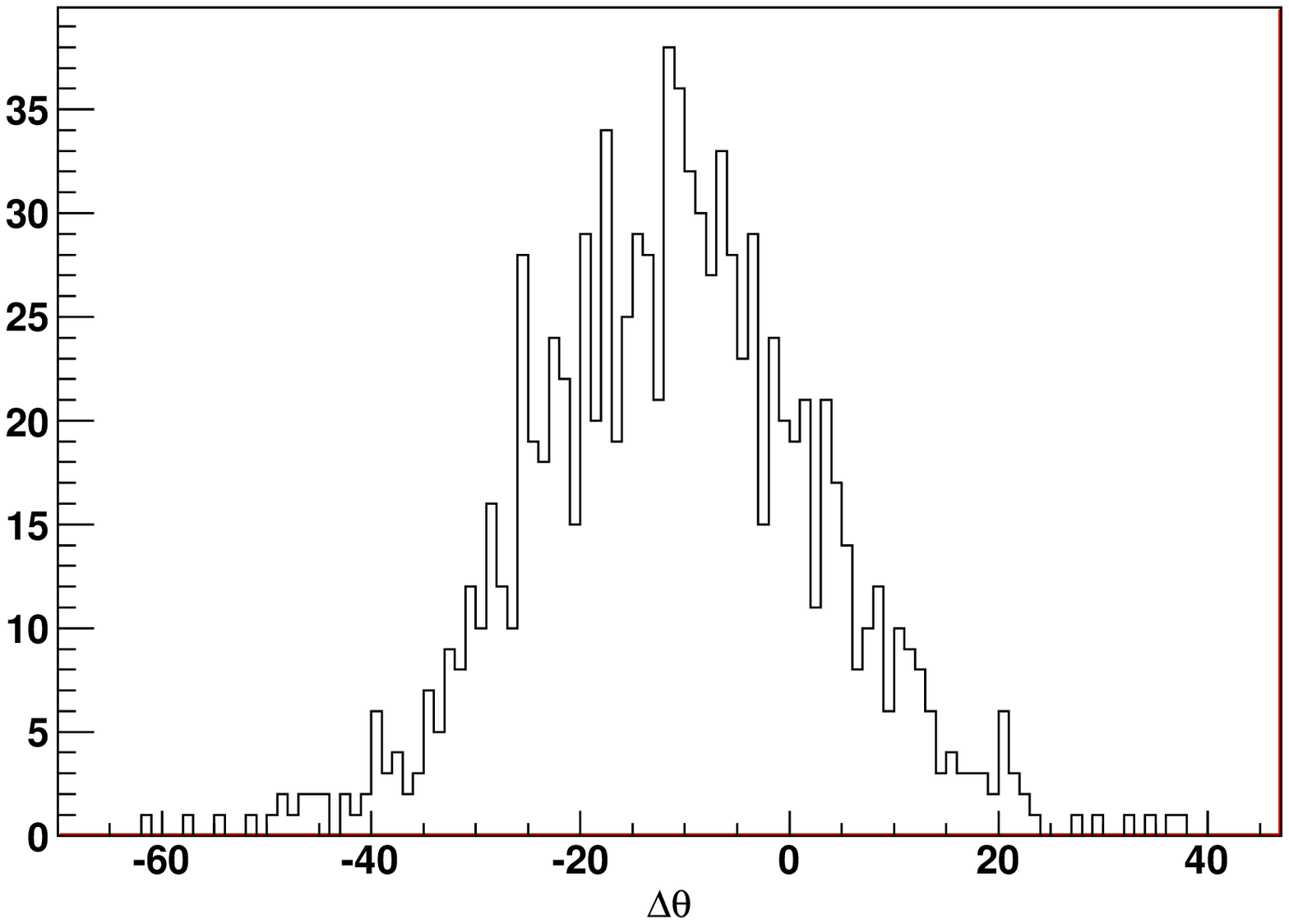}\\
\caption{The distribution of the bias factors, 
$k=\sqrt{C'_1 /C_1}$, $\Delta \theta(\theta^{'}-\theta)$ 
and $\Delta \phi(\phi^{'}-\phi)$ for number counts weighted by  
polarized flux after imposing the cuts
$S_{low} = 20$ mJy and $0.5<P<100$ mJy.} 
\label{fig:c1sim_PI}
\end{figure}

We show the distributions of bias factors, $k$, $\Delta\phi$ and
$\Delta\theta$, 
extracted from the 1000 simulated data sets,  
in Figs. \ref{fig:c1sim2_PI} and \ref{fig:c1sim_PI}. 
As the simulation is done for 
1000 realizations, the error in the mean values of
 $k$, $\Delta\phi$ and $\Delta\theta$ is suppressed by a factor of $\sqrt{1000}\sim31$, 
and negligible as compared to individual statistical error 
in dipole estimation.
The extracted values of the bias factors are given in Tables
\ref{tb:biasN}-\ref{tb:biasDOP}.  
The real map dipole parameters are computed after correcting for this bias.

The significance of dipole is estimated by comparing 
the bias corrected power, $C_1$, of the real data with the power, 
$\tilde{C}_1$, of 10,000 full sky 
isotropic random realizations.  
These random samples are generated by using the same procedure employed
for filling the empty pixels in real data, as described
above.  
The significance is quoted in terms of the P-value, which is defined as the probability
that an isotropic random sample may yield the dipole power larger than
that observed in real 
data. We estimate it by counting the number of random samples which yield
a higher power compared to real data. In case we find that none of 
the random power exceeds the real data, the P-value is less than $10^{-4}$,
which represents roughly a $4\sigma$ detection of the dipole anisotropy.

\begin{table}
\begin{tabular}{|c|c|c|c|c|}
\hline
{$S_{\text{low}}$} & {No. of Sources } & 
 $k$ & $\Delta\theta$ & $\Delta\phi$ \tabularnewline
\hline
\multicolumn{5}{|c|}{Polarization flux density  $0.5<P<100$ mJy} \\      
\hline
10 &   236864 &  $0.68\ (0.14)$ &  $ 11.2\ (12.6)$  &  $14.7\ (15.7)$ \\ 
20 &   159533 &  $0.72\ (0.16)$ &  $ 4.3\ (13.6)$  &  $13.0\ (15.0)$ \\ 
30 &   118806 &  $0.76\ (0.20)$ &  $ 2.3\ (16.1)$  &  $12.6\ (17.8)$ \\ 
40 &    93425 &  $0.75\ (0.24)$ &  $ -10.6\ (22.5)$  &  $10.0\ (26.0)$ \\ 
50 &    76205 &  $0.71\ (0.23)$ &  $ -19.8\ (23.6)$  &  $8.7\ (28.8)$ \\ 
75 &    50576 &  $0.70\ (0.23)$ &  $ -23.1\ (24.5)$  &  $8.4\ (34.8)$ \\ 
\hline
\multicolumn{5}{|c|}{degree of polarization $0.01<p<1.0$} \\      
\hline
10 & 280421 &  $0.53\ (0.09)$ & $ 12.8\ (9.6)$ & $16.3\ (13.3)$\\
20 & 162350 &  $0.65\ (0.13)$ & $ 10.5\ (11.5)$ & $13.9\ (14.6)$\\
30 & 110996 &  $0.70\ (0.16)$ & $ 9.2\ (13.8)$ & $13.5\ (15.9)$\\
40 &  82457 &  $0.75\ (0.22)$ & $ -1.6\ (19.8)$ & $12.1\ (21.7)$\\
50 &  64611 &  $0.77\ (0.23)$ & $ -11.2\ (19.5)$ & $10.2\ (23.8)$\\
75 &  40300 &  $0.77\ (0.24)$ & $ -10.7\ (19.5)$ & $9.6\ (24.7)$\\
\hline
\end{tabular}
\caption{The mean values of the bias factors $k$, $\Delta \theta$ and 
$\Delta\phi$ extracted from
simulations corresponding to number counts of significantly polarized sources 
with a cut on polarization 
flux density ($0.5<P<100$ mJy) and degree of
 polarization ($0.01  <p<1.0$).
These values correspond to the bias generated in the dipole
amplitude and direction due to the filling of masked sky with randomly generated
data. The values in brackets are the standard deviations over 1000 samples.
 }
\label{tb:biasN}
\end{table}

\begin{table}
\begin{tabular}{|c|c|c|c|c|}
\hline
{$S_{\text{low}}$} & {No. of Sources } & 
 $k$ & $\Delta\theta$ & $\Delta\phi$ \tabularnewline
\hline
10 &   236864 &  $0.70\ (0.15)$ &  $ -9.8\ (12.6)$  &  $9.9\ (12.5)$   
 \\
20 &   159533 &  $0.67\ (0.15)$ &  $ -11.0\ (14.0)$  &  $10.2\ (14.3)$   
 \\
30 &   118806 &  $0.68\ (0.17)$ &  $ -13.9\ (15.7)$  &  $9.3\ (17.3)$\\ 
40 &    93425 &  $0.63\ (0.16)$ &  $ -15.4\ (16.2)$  &  $8.5\ (19.7)$ \\
50 &    76205 &  $0.62\ (0.17)$ &  $-19.2\ (17.8)$  &  $8.7\ (21.7)$\\
75 &    50576 &  $0.65\ (0.18)$ &  $-18.1\ (18.7)$  &  $10.4\ (22.5)$\\
\hline
\end{tabular}
\caption{The mean values of the bias factors, $k$, $\Delta \theta$ and $\Delta\phi$ extracted from
simulations corresponding to number counts weighted by 
polarized flux density $P$ using the cut, $0.5<P<100$ mJy.
 The values in brackets are the standard deviations over 1000 samples.
 }
\label{tb:biasPI}
\end{table}

\begin{table}
\begin{tabular}{|c|c|c|c|c|}
\hline
{$S_{\text{low}}$} & {No. of Sources } 
 & $k$ & $\Delta\theta$ & $\Delta\phi$ 
\tabularnewline
\hline
10 & 280421 &  $0.67\ (0.10)$ & $ 8.0\ (8.2)$ & $13.8\ (9.3)$\\
20 & 162350 &  $0.66\ (0.11)$ & $ -0.6\ (9.0)$ & $11.0\ (9.5)$\\
30 & 110996 &  $0.67\ (0.14)$ & $ -5.0\ (11.9)$ & $10.1\ (12.4)$\\
40 &  82457 &  $0.61\ (0.14)$ & $ -13.4\ (14.1)$ & $6.8\ (15.6)$\\
50 &  64611 &  $0.57\ (0.14)$ & $ -17.1\ (14.3)$ & $9.2\ (17.6)$\\
75 &  40300 &  $0.58\ (0.15)$ & $ -19.2\ (15.6)$ & $5.7\ (19.6)$\\
\hline
\end{tabular}
\caption{The mean values of the bias factors, 
$k$, $\Delta \theta$ and $\Delta\phi$ extracted from
simulations corresponding to number counts weighted by
 degree of polarization $p$ ($0.01<p<1.0$). 
 The values in brackets are the standard deviations over 1000 samples.
 }
\label{tb:biasDOP}
\end{table}

\begin{figure}
\includegraphics[width=3.5in,angle=0]{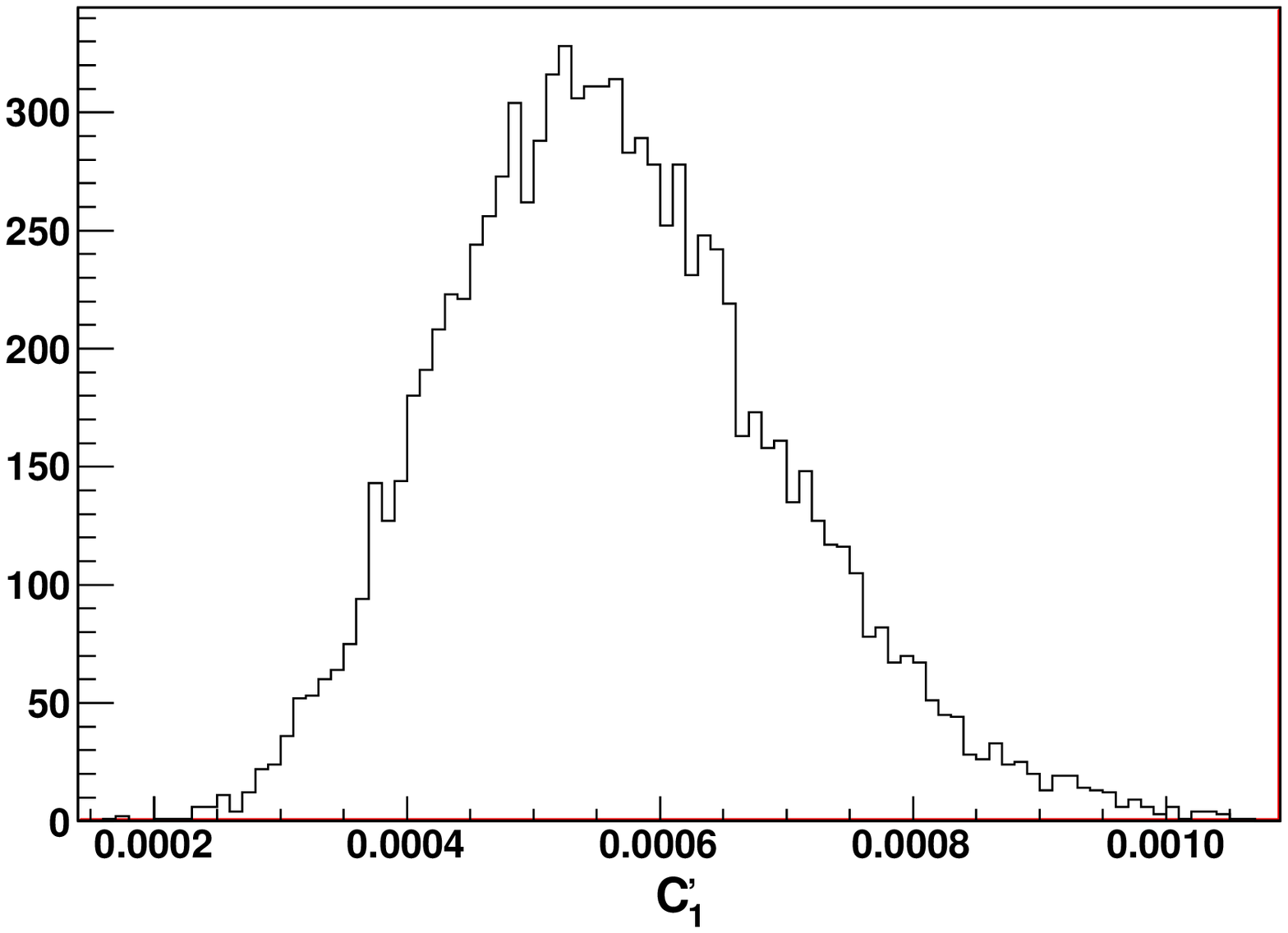}\\
\includegraphics[width=3.5in,angle=0]{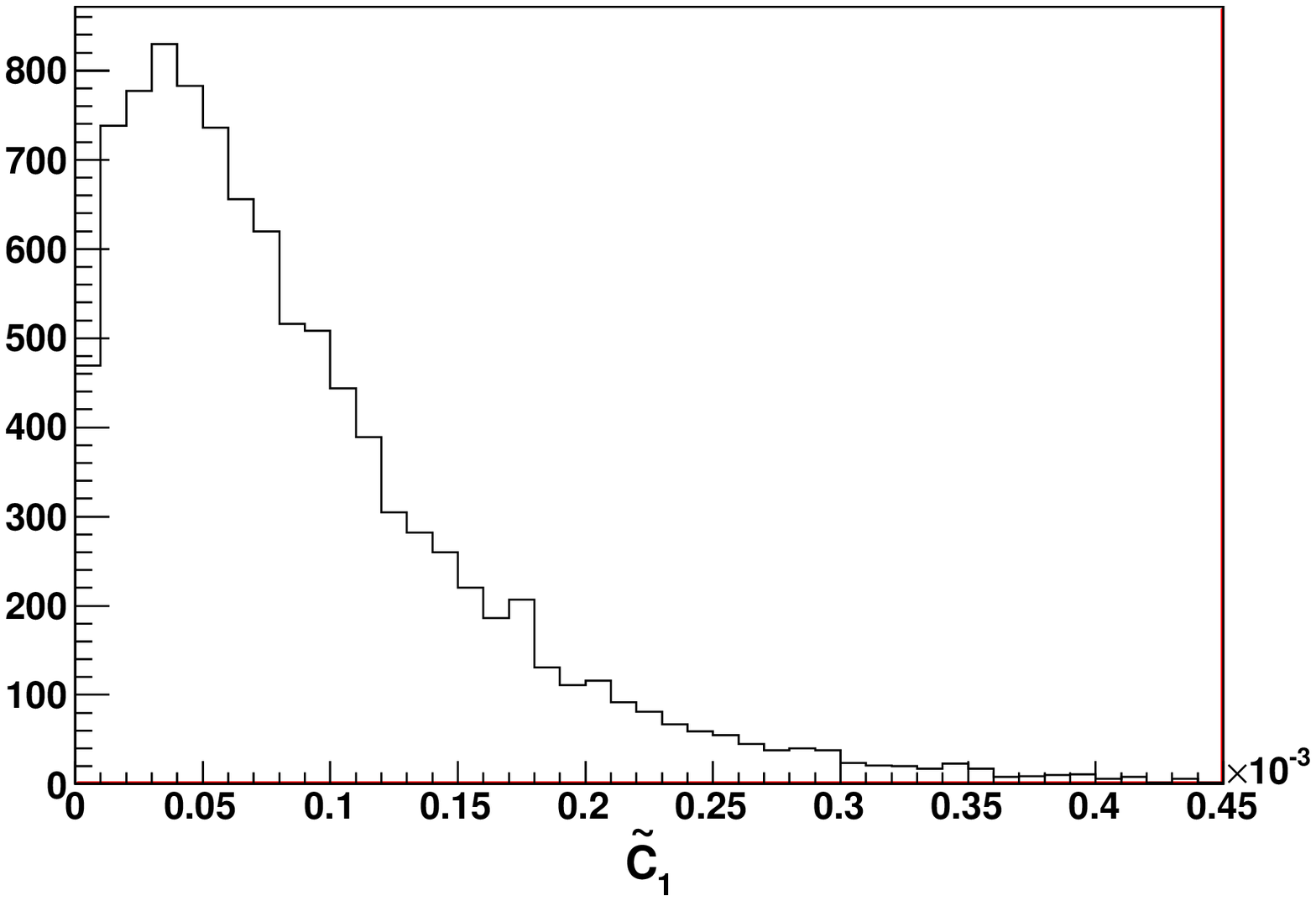}\\
\caption{The distribution of dipole power
for number counts of significantly polarized sources for the case
of real data (upper graph) and isotropic random simulated data (lower graph).  
 Here we have imposed the cuts $S_{low}=20$ mJy and $0.5<P<100$ mJy. 
The distribution in real data is obtained by randomly filling in the masked
regions, as explained in text.
}
\label{fig:c1N}
\end{figure}

\begin{table}
\begin{tabular}{|c|c|c|c|c|c|}
\hline
\multirow{2}{*}{$S_{\text{low}}$} & \multirow{2}{*}{$C'_{1}\left(\times10^{3}\right)$} & \multirow{2}{*}{$\tilde{C}_{1}\left(\times10^{3}\right)$} & \multirow{2}{*}{P-value} & \multirow{2}{*}{$\theta'$(\textdegree)} & \multirow{2}{*}{$\phi'$(\textdegree)}\tabularnewline
 &  &  &  &  & \tabularnewline
\hline
\multicolumn{6}{|c|}{polarized flux density  $0.5<P<100$ mJy } \\                                                            
\hline
10  &  $0.496\pm0.114$  &   $0.072\pm0.059$  &   $<10^{-4}$  &  $ 63\pm 9$  &  $151\pm 8$ \\ 
20  &  $0.565\pm0.133$  &   $0.090\pm0.074$  &   $< 10^{-4}$  &  $ 74\pm10$  &  $150\pm 8$ \\ 
30  &  $0.488\pm0.132$  &   $0.109\pm0.089$  &   $< 10^{-3}$  &  $ 77\pm12$  &  $152\pm 9$ \\ 
40  &  $0.351\pm0.135$  &   $0.134\pm0.110$  &   $2.3\times 10^{-3}$  &  $95\pm15$  &  $147\pm12$ \\ 
50  &  $0.440\pm0.165$  &   $0.161\pm0.132$  &   $9\times 10^{-4}$  &  $106\pm14$  &  $159\pm13$ \\ 
75  &  $0.577\pm0.227$  &   $0.220\pm0.182$  &   $1.6\times 10^{-3}$  &  $109\pm14$  &  $162\pm14$ \\ 
\hline
\multicolumn{6}{|c|}{degree of polarization $0.01<p<1$ } \\                                                                                          
\hline
10  &  $0.77\pm0.16$  &   $0.062\pm0.050$  &  $<10^{-4}$  &  $ 49\pm 7$  & $136\pm 6$ \\ 
20  &  $0.68\pm0.15$  &   $0.086\pm0.071$  &   $<10^{-4}$  &  $ 65\pm 9$  & $146\pm 7$ \\ 
30  &  $0.67\pm0.16$  &   $0.114\pm0.093$  &   $< 10^{-4}$  &  $ 69\pm10$  & $150\pm 8$ \\ 
40  &  $0.51\pm0.17$  &   $0.151\pm0.123$  &   $7\times 10^{-4}$  &  $ 84\pm14$  & $146\pm10$ \\ 
50  &  $0.62\pm0.20$  &   $0.188\pm0.154$  &   $8\times 10^{-4}$  &  $95\pm13$  & $158\pm11$ \\ 
75  &  $0.87\pm0.29$  &   $0.279\pm0.231$  &   $10\times 10^{-3}$  &  $95\pm14$  & $157\pm12$ \\ 
\hline
\end{tabular}
\caption{The extracted value of the dipole power $C'_1$ and the corresponding
value for simulated random data $\tilde C_1$ using number counts for
different cuts on flux density of a source ($S>S_{low}$) along with cut
on polarized flux density 0.5 mJy $<P<$ 100 mJy or  degree of polarization, $0.01<p<1$.
The significance of the dipole anisotropy is given in terms of the P-value.
The extracted parameters of the dipole axis,
$\theta'$ and $\phi'$ are also shown. 
}
\label{tb:C1N}
\end{table}

\section{Results}
\label{sc:PI_res}
\begin{figure}
\includegraphics[width=3.5in,angle=0]{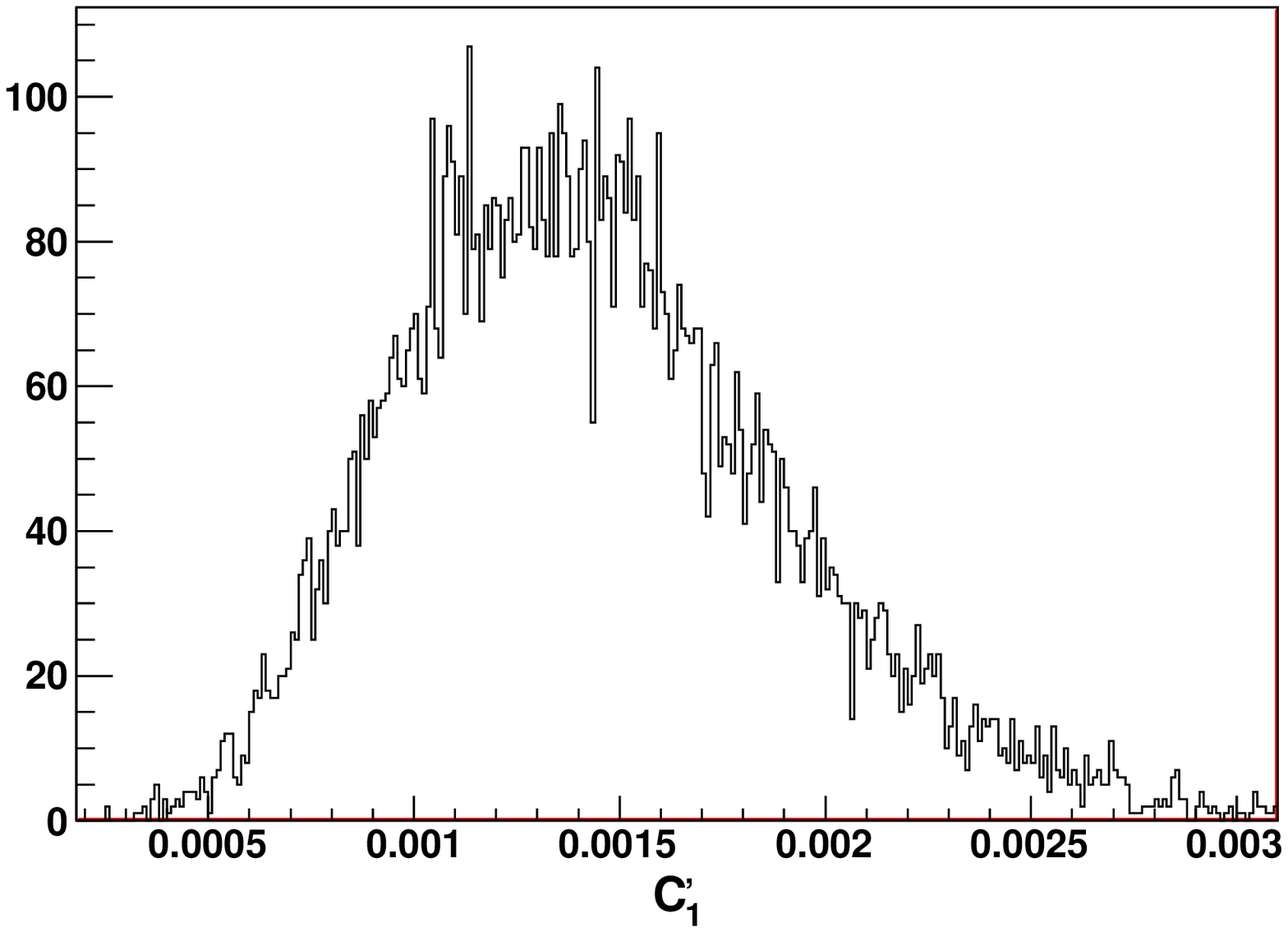}\\
\includegraphics[width=3.5in,angle=0]{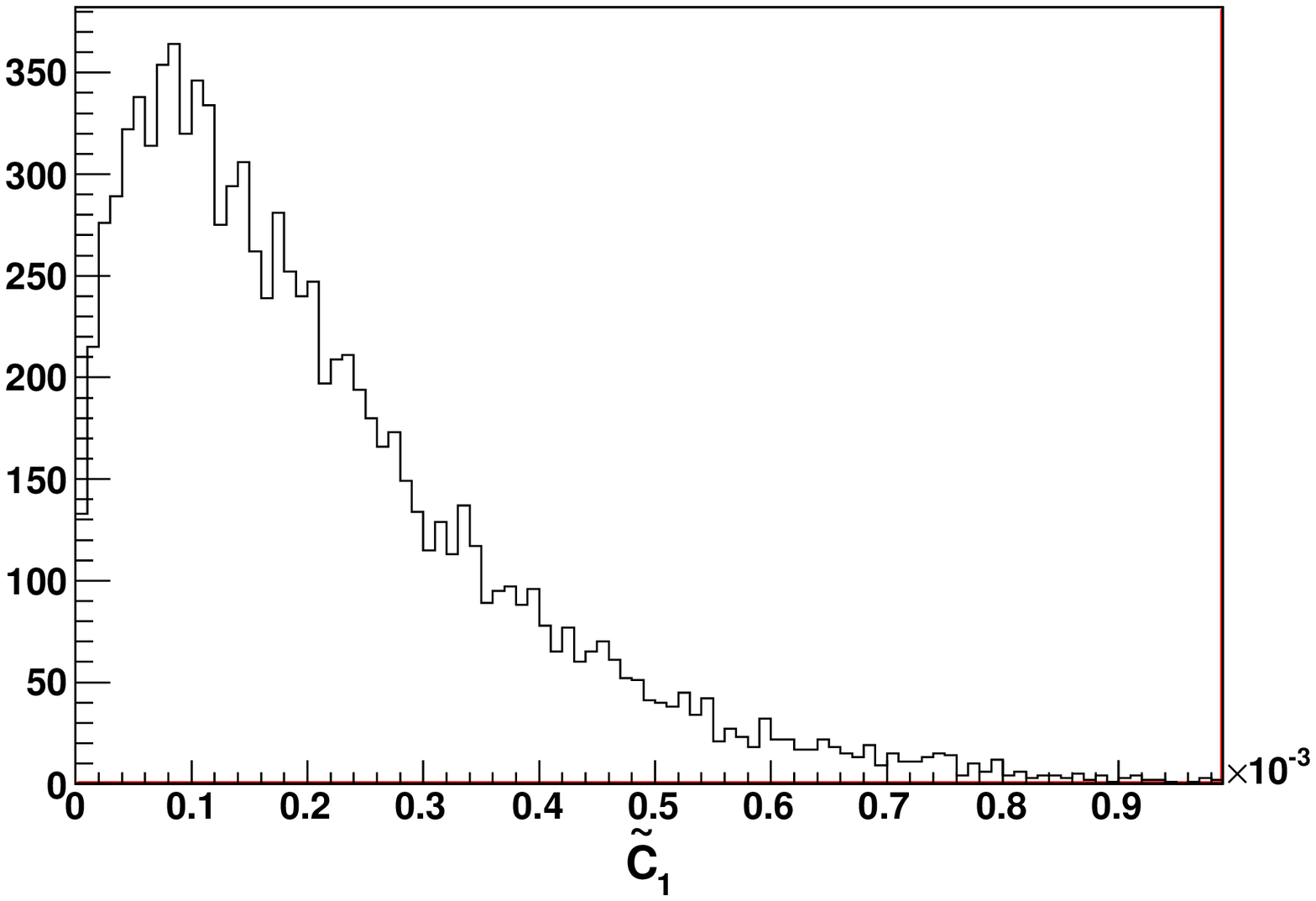}\\
\caption{The distribution of dipole power for number counts weighted by 
polarized flux for the case
of real data (upper graph) and isotropic random simulated data (lower graph). 
Here we have imposed the cuts $S_{low}=20$ mJy and $0.5<P<100$ mJy. 
The distribution in real data is obtained by randomly filling in the masked
regions, as explained in text.
}
\label{fig:c1PI}
\end{figure}

The distributions of $C'_1$ (real data) and $\tilde C_1$ (random
isotropic data) for number counts of significantly polarized sources
 after imposing the cuts $0.5<P<100$ mJy and $S_{low}=20$ mJy 
are shown in
Fig. \ref{fig:c1N}. The corresponding plots for number counts weighted
by polarized flux density are shown in Fig. \ref{fig:c1PI}. 
The distribution of real data corresponds to random filling of the
masked regions. The random map dipole power 
($\tilde{C}_1$) is used to calculate the significance of 
dipole.  
The results for 
extracted dipole power, $C'_1$, its significance and the direction parameters 
for number counts with a cut on polarized flux density, $0.5<P<100$ mJy, 
are given
 in Table \ref{tb:C1N} for different values of the lower limit
on the flux density,  $S_{low} =  10,20,30,40,50$ and 75  
mJy. The results with a 
  cut on degree of polarization, $0.01 <p< 1$,
are also shown.
The results for number counts weighted by
polarized flux are given in 
Table \ref{tb:C1PI}. The corresponding results for 
degree of polarization are given in Table \ref{tb:C1DOP}. 
The power corresponding
to random isotropic samples is also shown.
The data indicates a very 
significant signal of dipole anisotropy. In most 
cases the significance is $4\sigma$ or higher. The
significance increases as we reduce the limiting value $S_{low}$. This arises mainly
because of the increase in the number of sources as we reduce $S_{low}$, which
leads to a decrease in the dipole power, $\tilde C_1$, of isotropic random samples.

The bias corrected dipole amplitude and direction parameters for 
number counts, $D_N$ and $D_{Np}$, are given
in Tables \ref{tb:DN} and \ref{tb:DNdop} respectively. 
The results for number counts weighted
polarized flux density  
are given in Table \ref{tb:DPI}. 
The corresponding results for degree of polarization are 
 in Table \ref{tb:DDOP}. 
If we assume that the dipole arises primarily due to our local motion,
the speed turns out to be much larger in comparison to the 
CMBR expectations. The exponents $x$ for the power law fit to 
the number density $n(S,P)$ are found to be, $x=0.759,0.885,0.959,1.02,1.06, 1.25$
for the cuts
 $S_{low}=$ 10, 20, 30,40, 50 and 75 mJy respectively. 
This is obtained by imposing no cut on $P$. The exponent $x_P$ is determined
from the data after imposing the cut $0.5 < P< 100$ mJy and no cut on $S$.
We obtain $x_P=1.2$. 
As discussed in section
\ref{sc:analyses},
we set $\alpha_P =0.75$.    
The extracted speed for the case of number counts weighted by
polarized flux density is given in  
Table \ref{tb:speed1}. 
Here the speed is extracted assuming that 
 the dipole arises entirely due to kinematic effects. 
We find that the extracted dipole cannot be consistently attributed to 
kinematic effects, predicted on the basis of CMBR dipole. The 
speed turns out to be about 6-7 times larger.   
Similar results are obtained with number counts.  
For comparison, the extracted dipole from number counts
as well as sky brightness, without considering polarization properties
of the radio sources, is found to be about 3-4 times larger than
that predicted by CMBR  
 \citep{Kothari:2013}. 
This indicates the presence of a larger
intrinsic dipole in the polarized sample. 
Hence we speculate that the physical mechanism responsible for 
the intrinsic dipole affects the polarization more than the flux. 

We find that the polar angle, $\theta$, of the dipole axis 
shows a strong dependence on $S_{low}$ for several cases. In contrast,
the dipole amplitude as well as the azimuthal angle are found to be relatively
stable. The dependence on $S_{low}$ can only be attributed to the 
presence of $\theta$ dependent bias in the number counts
of significantly polarized sources. Such a bias is 
also present in the complete data set, which includes both unpolarized
and polarized sources \citep{Blake:2002}. In this case it practically
gets eliminated in data with $S>15$ mJy. However in the present case
the bias seems to be present for larger values of the lower limit 
on the flux density. We notice that the   
parameters for the case of polarized flux weighted number counts,
Table \ref{tb:DPI},
are relatively stable 
with the cut on flux density, $S$. 
This trend 
is understandable since, in this case, the contribution
due to the data with low flux is reduced. This reduces the systematic
effects that might be present due to faint sources. Similar trend was
seen for the case of flux density weighted number counts \citep{Kothari:2013}.
Hence the results for the case of number counts weighted by polarization
flux density may be relatively free of bias. For this reason
we study this observable in considerable detail.

\begin{figure}
\includegraphics[width=3.5in,angle=0]{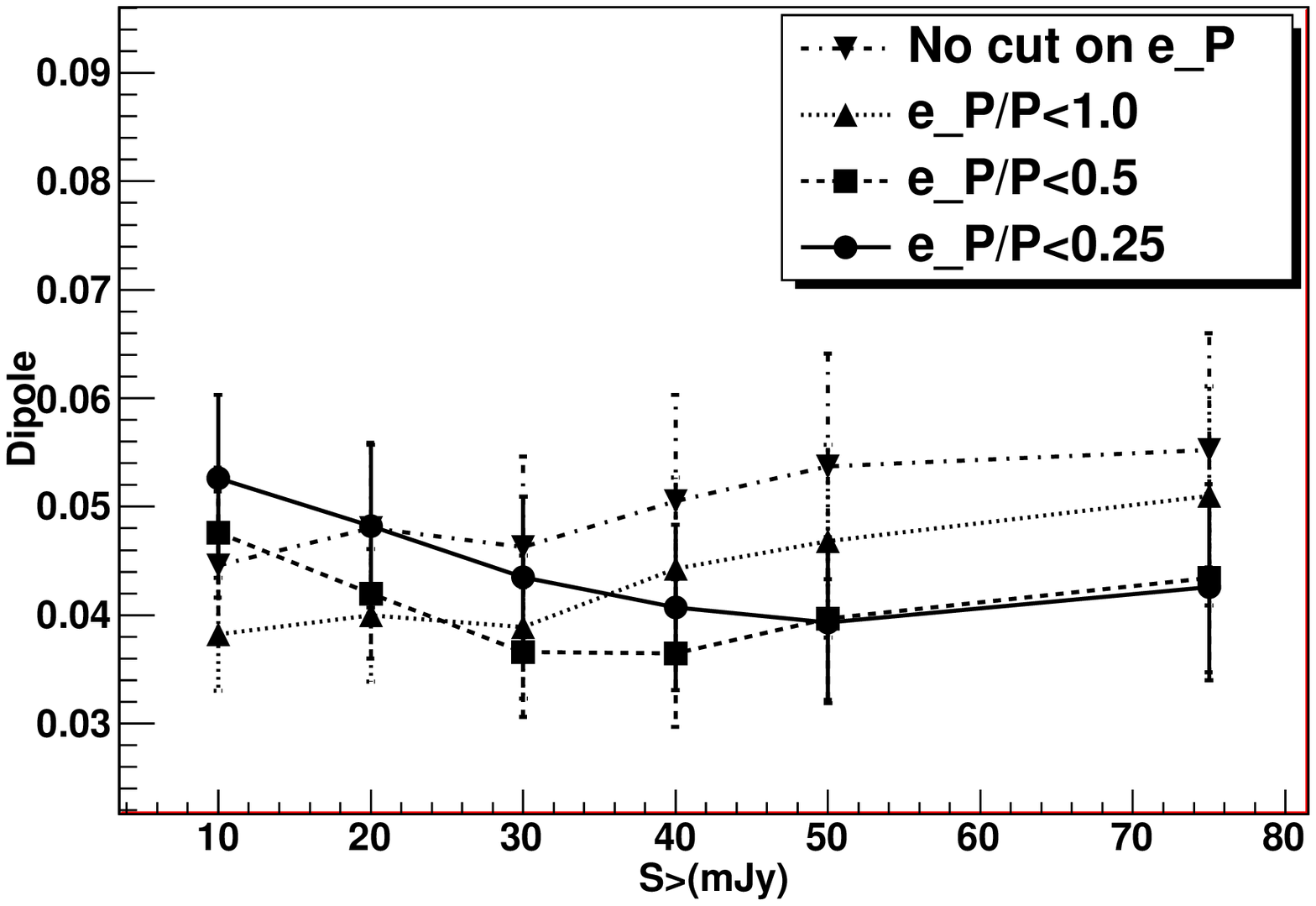}\\
\includegraphics[width=3.5in,angle=0]{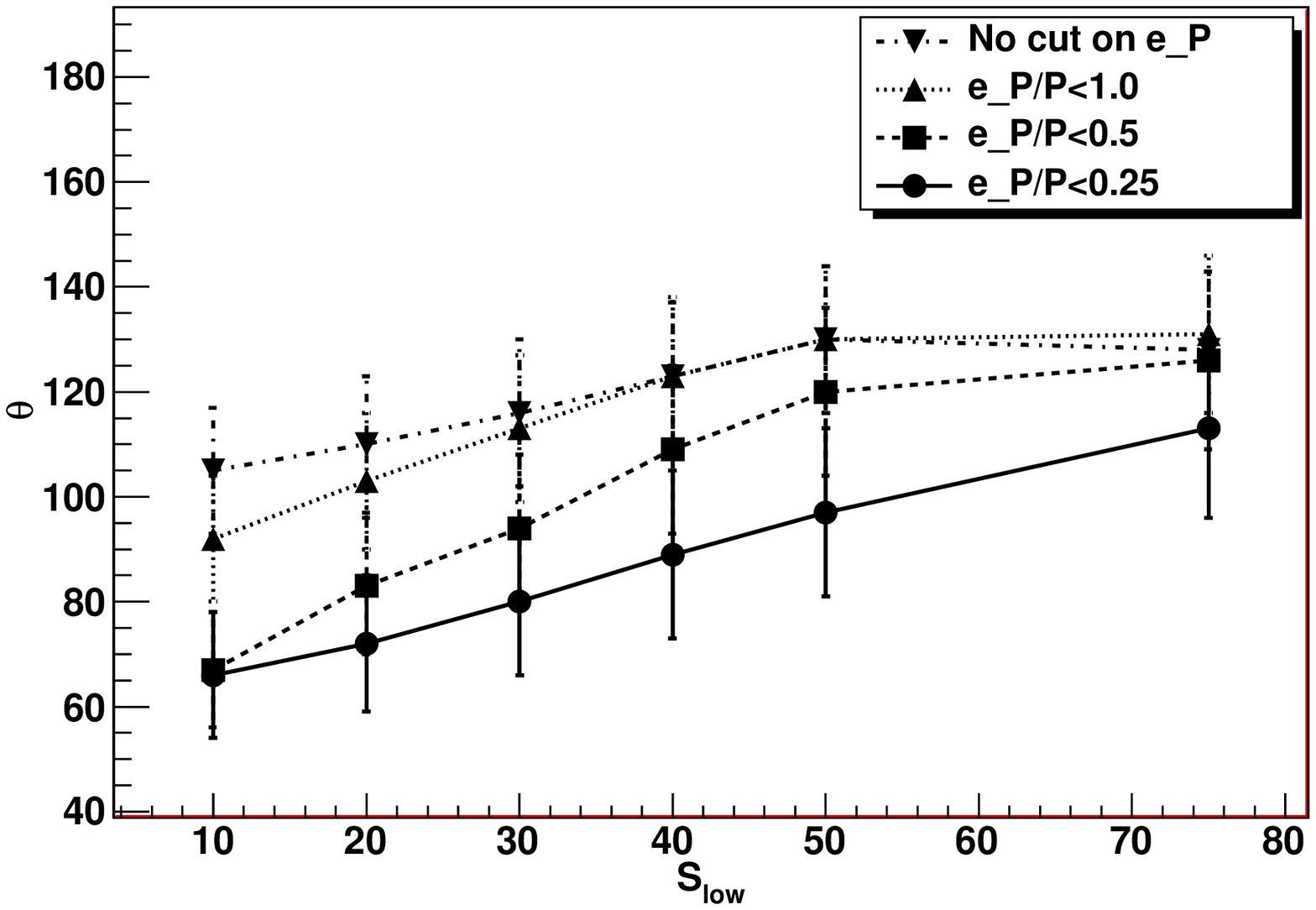}\\
\includegraphics[width=3.5in,angle=0]{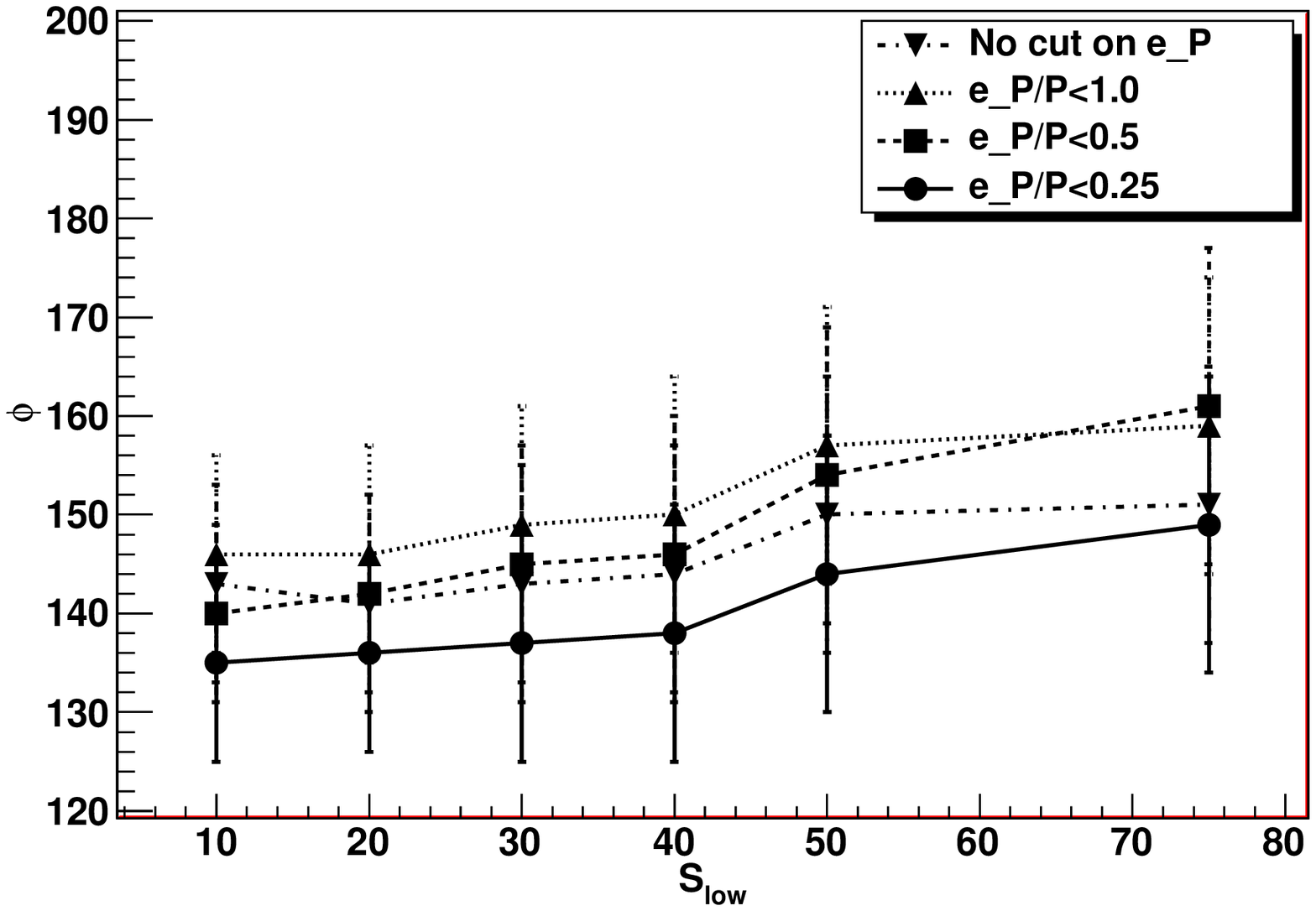}\\
\caption{The dipole amplitude (upper graph), polar angle, $\theta$, (middle 
graph) and $\phi$ or RA (lower graph) for  
number counts weighted by polarized flux as a function of lower limit
on the flux density, $S_{low}$ for different cuts on the fractional error in the
polarized flux, $e_P/P< 1, 0.5, 0.25$. Here we have used the additional  
  cut $0.5<P<100$ mJy.
}
\label{fig:dipole_flux}
\end{figure}

\begin{table}
\begin{tabular}{|c|c|c|c|c|c|}
\hline
\multirow{2}{*}{$S_{\text{low}}$} & \multirow{2}{*}{$C'_{1}\left(\times10^{3}\right)$} & \multirow{2}{*}{$\tilde{C}_{1}\left(\times10^{3}\right)$} & \multirow{2}{*}{P-value} & \multirow{2}{*}{$\theta'$(\textdegree)} & \multirow{2}{*}{$\phi'$(\textdegree)}\tabularnewline
 &  &  &  &  & \tabularnewline
\hline

10  &  $1.36\pm0.41$  &   $0.17\pm0.14$  &   $< 10^{-4}$  &  $95\pm12$  &  $153\pm10$ \\ 
20  &  $1.45\pm0.48$  &   $0.21\pm0.17$  &   $< 10^{-4}$  &  $99\pm13$  &  $151\pm11$ \\ 
30  &  $1.39\pm0.50$  &   $0.26\pm0.21$  &   $< 10^{-4}$  &  $103\pm14$  &  $152\pm12$ \\ 
40  &  $1.41\pm0.55$  &   $0.30\pm0.25$  &   $< 10^{-4}$  &  $108\pm14$  &  $152\pm13$ \\ 
50  &  $1.55\pm0.60$  &   $0.35\pm0.29$  &   $< 10^{-4}$  &  $111\pm14$  &  $159\pm14$ \\
75  &  $1.80\pm0.71$  &   $0.46\pm0.38$  &   $< 10^{-4}$  &  $110\pm15$  &  $162\pm14$ \\
\hline
\end{tabular}
\caption{The extracted value of the dipole power $C'_1$ and the corresponding
value for simulated isotropic data $\tilde C_1$ using number count weighted by
polarized flux density $P$ ($0.5<P<100$) 
for different cuts on flux density of a source ($S>S_{low}$).
The significance of the dipole anisotropy, P-value, as well as the
extracted dipole axis parameters,
$\theta'$ and $\phi'$ are also shown.
}
\label{tb:C1PI}
\end{table}

\begin{table}
\begin{tabular}{|c|c|c|c|c|c|}
\hline
\multirow{2}{*}{$S_{\text{low}}$} & \multirow{2}{*}{$C'_{1}\left(\times10^{3}\right)$} & \multirow{2}{*}{$\tilde{C}_{1}\left(\times10^{3}\right)$} & \multirow{2}{*}{P-value} & \multirow{2}{*}{$\theta'$(\textdegree)} & \multirow{2}{*}{$\phi'$(\textdegree)}\tabularnewline
 &  &  &  &  & \tabularnewline
\hline
10  &  $1.32\pm0.31$  & $0.078\pm0.064$  &  $<10^{-4}$  &  $ 70\pm 10$  & $149\pm 8$ \\ 
20  &  $1.57\pm0.44$  & $0.12\pm0.09$  &  $<10^{-4}$  &  $ 83\pm11$  &  $141\pm 8$ \\ 
30  &  $1.26\pm0.42$  & $0.16\pm0.13$  &   $<10^{-4}$  &  $ 89\pm13$  &  $142\pm 10$ \\ 
40  &  $1.18\pm0.47$  & $0.21\pm0.17$  &   $< 10^{-4}$  &  $103\pm14$  &  $139\pm11$ \\ 
50  &  $1.37\pm0.54$  & $0.26\pm0.21$  &   $< 10^{-4}$   &  $115\pm13$  &  $154\pm13$ \\ 
75  &  $1.76\pm0.68$  & $0.38\pm0.31$  &   $< 10^{-4}$   &  $116\pm13$  &  $160\pm14$ \\ 
\hline
\hline
\end{tabular}
\caption{The extracted value of the dipole power $C'_1$ and the corresponding
value for simulated isotropic data $\tilde C_1$ using number counts
weighted by the degree of polarization $p$ ($0.01<P<1$) for different cuts on flux density of a source ($S>S_{low}$).
The significance of the dipole anisotropy, P-value, as well as the
extracted dipole axis parameters,
$\theta'$ and $\phi'$, are also shown.
}
\label{tb:C1DOP}
\end{table}

\begin{table}
\begin{tabular}{|c|c|c|c|}
\hline
\multirow{2}{*}{$S_{\text{low}}$} & \multirow{2}{*}{$D_N$} & \multirow{2}{*}{$\theta$(\textdegree)} & \multirow{2}{*}{$\phi$(\textdegree)} \\
&  & &   \tabularnewline
\hline
10 &  $0.028\pm0.003$      &  $ 51\pm9$  &   $137\pm 8$ \\ 
20 &  $0.028\pm0.003$      &  $ 70\pm10$  &   $137\pm 8$ \\ 
30 &  $0.025\pm0.003$      &  $ 75\pm12$  &   $140\pm 9$ \\ 
40 &  $0.021\pm0.004$      &  $106\pm15$  &   $137\pm12$ \\ 
50 &  $0.025\pm0.005$      &  $126\pm14$  &   $151\pm13$ \\
75 &  $0.029\pm0.006$      &  $132\pm14$  &   $154\pm14$ \\
\hline
\end{tabular}
\hfill
\caption{The extracted dipole amplitude for number counts, $D_N$, and 
the corresponding direction parameters ($\theta$, $\phi$) for various 
cuts on the flux density for the cut on polarized flux,
$0.5<P<100$ mJy. }
\label{tb:DN}
\end{table}

\begin{table}
\begin{tabular}{|c|c|c|c|}
\hline
\multirow{2}{*}{$S_{\text{low}}$} & \multirow{2}{*}{$D_{Np}$} & \multirow{2}{*}{$\theta$(\textdegree)} & \multirow{2}{*}{$\phi$(\textdegree)} \\
&  & &   \tabularnewline
\hline
10 &  $0.044\pm0.005$ &   $ 36\pm 7$  &   $120\pm 6$ \\ 
20 &  $0.034\pm0.004$ &   $ 55\pm 9$  &   $132\pm 7$ \\ 
30 &  $0.031\pm0.004$ &   $ 60\pm10$  &   $136\pm 8$ \\ 
40 &  $0.026\pm0.004$ &   $86\pm14$  &   $134\pm10$ \\ 
50 &  $0.027\pm0.004$ &   $106\pm13$  &   $148\pm11$ \\ 
75 &  $0.033\pm0.005$ &   $106\pm14$  &   $147\pm12$ \\ 
\hline
\end{tabular}
\hfill
\caption{The extracted dipole amplitude for number counts, $D_{Np}$, and 
the corresponding direction parameters ($\theta$, $\phi$) for various 
cuts on the flux density for the cut on
degree of polarization, $0.01<p<1.0$.}
\label{tb:DNdop}
\end{table}

\begin{table}
\begin{tabular}{|c|c|c|c|}
\hline
\multirow{2}{*}{$S_{\text{low}}$} & \multirow{2}{*}{$D_P$} &  \multirow{2}{*}{$\theta$(\textdegree)} & \multirow{2}{*}{$\phi$(\textdegree)} \\
&  &  &  \tabularnewline
\hline
10 &  $0.045\pm0.007$  &    $105\pm12$  &  $143\pm10$  \\ 
20 &  $0.048\pm0.008$  &    $110\pm13$  &  $141\pm11$  \\ 
30 &  $0.046\pm0.008$  &    $116\pm14$  &  $143\pm12$  \\ 
40 &  $0.051\pm0.010$  &    $123\pm14$  &  $144\pm13$  \\ 
50 &  $0.054\pm0.010$  &   $130\pm14$  &  $150\pm14$  \\
75 &  $0.055\pm0.011$  &   $128\pm15$  &  $151\pm14$  \\
\hline      
\end{tabular}
\hfill
\caption{The extracted dipole amplitude, $D_{P}$, for number counts weighted
by the polarized flux density and the corresponding
direction parameters ($\theta$, $\phi$) as a function of $S_{low}$ for the
polarized flux density lying in the range $0.5<P<100$ mJy.
 }
\label{tb:DPI}
\end{table}

\begin{table}
\begin{tabular}{|c|c|c|c|}
\hline
\multirow{2}{*}{$S_{\text{low}}$} & \multirow{2}{*}{$D_p$} &  \multirow{2}{*}{$\theta$(\textdegree)} & \multirow{2}{*}{$\phi$(\textdegree)} \\
  &  &  &\tabularnewline
\hline
10 &  $0.046\pm0.005$  &    $ 62\pm10$  &  $135\pm 8$  \\ 
20 &  $0.051\pm0.007$  &    $ 84\pm11$  &  $130\pm 8$  \\ 
30 &  $0.045\pm0.008$  &   $94\pm13$  &  $132\pm10$  \\ 
40 &  $0.048\pm0.010$  &    $117\pm14$  &  $132\pm11$  \\ 
50 &  $0.055\pm0.011$  &   $132\pm13$  &  $144\pm13$  \\ 
75 &  $0.061\pm0.012$  &   $136\pm13$  &  $154\pm14$  \\ 
\hline      
\end{tabular}
\hfill
\caption{The extracted dipole amplitude, $D_p$, for the number counts weighted
by the degree of polarization and the corresponding
direction parameters ($\theta$, $\phi$) for various cuts on the flux density.
Here we have imposed the cut on degree of polarization such that, $0.01<p<1$. 
} 
\label{tb:DDOP}
\end{table}

\begin{table}
\begin{tabular}{|c|c|c|c|c|c|c|}
\hline
$S_{low}$ & 10 &20 &30 &40 &50&75\\
\hline
 $v$ (Km/s) & $2490\pm390$    
  &  $2550\pm420$   
 &  $2390\pm420$    
 &  $2600\pm510$    
 &  $2720\pm500$    
 &  $2620\pm520$   \\ 
\hline
\end{tabular}
\caption{The extracted local speed for number counts weighted by
polarized flux density, with the cut
 $0.5<P<100$ mJy.}
\label{tb:speed1}
\end{table}

\subsection{The super-galactic plane} 
In our analysis we have eliminated sources which are likely to belong
to the local supercluster using the catalogues of nearby galaxies
\citep{Saunders:2000,Vaucouleurs:1991,Corwin:1994}. Here we follow an alternate procedure
to eliminate the bias due to such sources. Their density  
is largest near the supergalactic plane. Hence we can 
determine their influence by masking this plane. 
Here we eliminate  
  the region lying within $\pm 10^o$
of the supergalactic plane, besides imposing other cuts, described earlier. 
Here we study only the dipole amplitude, $D_P$, corresponding to the
number counts weighted by polarized flux density. As we have argued
above, this observable appears to be relatively free of bias.
The extracted values of 
bias corrected dipole amplitude and
direction parameters are given in  
Table \ref{tb:DPI_SupGal}. We find that the amplitude does not change
much with the super-galactic cut.
 Hence we conclude that the 
dipole anisotropy does not get significant contribution due to the local 
clustering effect. It is most likely a cosmological effect.
The largest change for polarized flux weighted number counts 
is seen in the $\phi$ (or RA). However in this case also
the results with and without the supergalactic cut agree within errors.

\begin{table}
\begin{tabular}{|c|c|c|c|}
\hline
\multirow{2}{*}{$S_{\text{low}}$} & \multirow{2}{*}{$D_P$} &  \multirow{2}{*}{$\theta$(\textdegree)} & \multirow{2}{*}{$\phi$(\textdegree)} \\
&  &  &  \tabularnewline
\hline      

10 &  $0.044\pm0.010$  &  $112\pm17$  &  $139\pm15$  \\ 
20 &  $0.048\pm0.011$  &  $118\pm18$  &  $136\pm16$  \\ 
30 &  $0.048\pm0.012$  &  $121\pm19$  &  $138\pm17$  \\ 
40 &  $0.051\pm0.013$  &  $131\pm20$  &  $139\pm19$  \\ 
50 &  $0.054\pm0.014$  &  $135\pm20$  &  $146\pm21$  \\ 
75 &  $0.054\pm0.015$  &  $133\pm22$  &  $150\pm22$  \\ 

\hline      
\end{tabular}
\hfill
\caption{The extracted dipole amplitude $D_{P}$ and the corresponding
direction parameters ($\theta$, $\phi$) for various cuts on the flux density
$S$ and super galactic plane ($|b|>10^\circ$).
Here the polarized flux is confined to lie in the range $0.5<P<100$ mJy. }
\label{tb:DPI_SupGal}
\end{table}

\subsection{Polarized flux weighted number counts}
In this section we study the dipole anisotropy, $D_P$, in
 number counts weighted by polarized 
flux density, $P_I$, in more detail by imposing further cuts to
improve the quality of data. 
This observable is equal to the
total polarized flux density in any pixel. 
As discussed above, we expect it to
be more stable in comparison to pure number counts, which are dominated
by low flux sources. 
We study the dipole anisotropy, $D_P$, by imposing further cuts on
the fractional error in polarized flux, $(e_P/P)$. 
 We find that many sources, especially those with 
low polarized flux, have large error. Here we study how our results change
as we impose the cuts, $e_P/P<$ 1, 0.5 and 0.25.  
Furthermore we also determine the results with a more stringent cut
on the polarized flux, $1<P<100$ mJy.

In Tables \ref{tb:DPI_11}  and \ref{tb:DPI_12} we show the extracted 
dipole for a different cuts on the error in the polarized flux with a lower limit
on the polarized flux equal to 0.5 mJy and 1.0 mJy respectively. 
We see that after imposing 
a cut on the error in polarized flux density, $e_P/P<1$, the dipole amplitude 
becomes smaller in comparison to the case with no cut. However as
we impose more stringent cuts, the amplitude shows relatively small
change. One still observes some dependence on the lower limit of
the flux density $S_{low}$. However this tends to saturate as we 
go to larger values of $S_{low}$. 
In Fig. \ref{fig:dipole_flux}, we plot the dipole amplitude and the direction
parameters, $(\theta,\phi)$, as a function of the lower limit, $S_{low}$,
on the flux density. We find that the extracted values of
the dipole power and $\phi$ agree with one another within errors
 for different cuts. The values of $\theta$ show a mild increase with
$S_{low}$ which tends to saturate for large values of $S_{low}$. 
In Fig. \ref{fig:dipole_flux}, such a saturation is not observed for
the cut $e_P/P<0.25$. However for this case also the extracted value
of $\theta$ becomes uniform as we increase $S_{low}$ to values larger than
75 mJy. Hence we
do not observe a very strong dependence of any of these extracted parameters
to the different cuts imposed.  

Finally we determine how our results are affected by the bright and extended 
radio sources. Some of these sources would be misidentified as a cluster
of very large number of sources in the NVSS survey \citep{Blake:2002}. 
Hence these are likely to introduce bias in our analysis.
\cite{Blake:2002} identify 22 such regions. \cite{Kothari:2013}
show that after imposing a cut which removes such sources, the extracted
dipole amplitude gets reduced by about $10-20$ \%. In the present case
of significantly polarized sources, however, we find that the contribution
of these sources is much smaller. After imposing this cut, we find that
the dipole amplitude changes by less than 10\%. The direction parameters
are found to be even less sensitive. Hence we conclude that these 
sources do not significantly affect our results. 

\begin{table}
\begin{tabular}{|c|c|c|c|}
\hline
\multirow{2}{*}{$S_{\text{low}}$} & \multirow{2}{*}{$D_P$} &  \multirow{2}{*}{$\theta$(\textdegree)} & \multirow{2}{*}{$\phi$(\textdegree)} \\
&  &  &  \tabularnewline
\hline      
\multicolumn{4}{|c|}{$e_P/P < 1$}\\ 
\hline      

10 &  $0.038\pm0.005$  &  $92\pm12$  &  $146\pm10$  \\ 
20 &  $0.040\pm0.006$  &  $103\pm13$  &  $146\pm11$  \\ 
30 &  $0.039\pm0.007$  &  $113\pm14$  &  $149\pm12$  \\ 
40 &  $0.044\pm0.008$  &  $123\pm15$  &  $150\pm14$  \\ 
50 &  $0.047\pm0.009$  &  $130\pm14$  &  $157\pm14$  \\ 
75 &  $0.051\pm0.010$  &  $131\pm15$  &  $159\pm15$  \\ 

\hline      
\multicolumn{4}{|c|}{$e_P/P < 0.5$}\\ 
\hline      

10 &  $0.048\pm0.006$  &  $67\pm11$  &  $140\pm9$  \\ 
20 &  $0.042\pm0.006$  &  $83\pm13$  &  $142\pm10$  \\ 
30 &  $0.037\pm0.006$  &  $94\pm14$  &  $145\pm12$  \\ 
40 &  $0.037\pm0.007$  &  $109\pm16$  &  $146\pm14$  \\ 
50 &  $0.040\pm0.008$  &  $120\pm16$  &  $154\pm15$  \\ 
75 &  $0.043\pm0.009$  &  $126\pm17$  &  $161\pm16$  \\ 

\hline      
\multicolumn{4}{|c|}{$e_P/P < 0.25$}\\ 
\hline      

10 &  $0.053\pm0.008$  &  $66\pm12$  &  $135\pm10$  \\ 
20 &  $0.048\pm0.008$  &  $72\pm13$  &  $136\pm10$  \\ 
30 &  $0.044\pm0.007$  &  $80\pm14$  &  $137\pm12$  \\ 
40 &  $0.041\pm0.008$  &  $89\pm16$  &  $138\pm13$  \\ 
50 &  $0.040\pm0.007$  &  $97\pm16$  &  $144\pm14$  \\ 
75 &  $0.043\pm0.009$  &  $113\pm17$  &  $149\pm15$  \\ 
\hline      

\end{tabular}
\hfill
\caption{The extracted dipole amplitude $D_{P}$ and the corresponding
direction parameters ($\theta$, $\phi$) for various cuts on the flux density 
and the fractional error on the polarized flux. Here the 
polarized flux density lies in the range $0.5-100$ mJy. 
 }
\label{tb:DPI_11}
\end{table}

\begin{table}
\begin{tabular}{|c|c|c|c|}
\hline
\multirow{2}{*}{$S_{\text{low}}$} & \multirow{2}{*}{$D_P$} &  \multirow{2}{*}{$\theta$(\textdegree)} & \multirow{2}{*}{$\phi$(\textdegree)} \\
&  &  &  \tabularnewline
\hline      
\multicolumn{4}{|c|}{$e_P/P < 1$}\\ 
\hline      

10 &  $0.042\pm0.006$  &  $98\pm12$  &  $150\pm10$  \\ 
20 &  $0.042\pm0.006$  &  $105\pm13$  &  $146\pm11$  \\ 
30 &  $0.042\pm0.007$  &  $114\pm14$  &  $151\pm12$  \\ 
40 &  $0.044\pm0.008$  &  $123\pm15$  &  $150\pm14$  \\ 
50 &  $0.047\pm0.009$  &  $129\pm15$  &  $159\pm14$  \\ 
75 &  $0.049\pm0.010$  &  $129\pm15$  &  $160\pm15$  \\ 

\hline      
\multicolumn{4}{|c|}{$e_P/P < 0.5$}\\ 
\hline      

10 &  $0.042\pm0.005$  &  $82\pm12$  &  $145\pm10$  \\ 
20 &  $0.040\pm0.006$  &  $93\pm13$  &  $146\pm11$  \\ 
30 &  $0.037\pm0.006$  &  $104\pm15$  &  $149\pm12$  \\ 
40 &  $0.040\pm0.007$  &  $116\pm16$  &  $149\pm14$  \\ 
50 &  $0.042\pm0.008$  &  $125\pm16$  &  $157\pm15$  \\ 
75 &  $0.044\pm0.009$  &  $125\pm16$  &  $161\pm16$  \\ 

\hline      
\multicolumn{4}{|c|}{$e_P/P < 0.25$}\\ 
\hline      
10 &  $0.053\pm0.008$  &  $66\pm12$  &  $135\pm10$  \\ 
20 &  $0.048\pm0.008$  &  $71\pm13$  &  $136\pm10$  \\ 
30 &  $0.044\pm0.007$  &  $79\pm14$  &  $137\pm12$  \\ 
40 &  $0.041\pm0.008$  &  $88\pm16$  &  $138\pm13$  \\ 
50 &  $0.039\pm0.007$  &  $97\pm16$  &  $144\pm14$  \\ 
75 &  $0.043\pm0.009$  &  $113\pm17$  &  $149\pm15$  \\

\hline      
\end{tabular}
\hfill
\caption{The extracted dipole amplitude $D_{P}$ and the corresponding
direction parameters ($\theta$, $\phi$) for various cuts on the flux density
and the fractional error on polarized flux.
Here the polarized flux density lies in the range $1.0-100$ mJy. 
 }
\label{tb:DPI_12}
\end{table}

\section{Discussion and Conclusion}
\label{sc:physics}
In this paper we have studied the dipole anisotropy in the
number counts of radio sources after imposing cuts on the polarized flux
density and independently on the degree of polarization. We have
also studied the anisotropy in the number counts weighted by
polarized flux and by degree of polarization. This study is important
in view of the recently reported dipole anisotropy in radio source
distribution \citep{Blake:2002,Singal:2011,Gibelyou:2012,Rubart:2013,Kothari:2013}. Most of these studies find a dipole amplitude much larger than that
predicted on the basis of CMBR dipole.  
We find that the dipole amplitude gets further enhanced after imposing 
cuts on polarized flux
and the degree of polarization. 
The amplitude is found to be more than twice as large as that observed
in the case unpolarized observables. The direction of the dipole is again 
found to be relatively close to the CMBR dipole. 
The dipole parameters, however, are found to show considerable
dependence on the lower limit, $S_{low}$, imposed on the flux density. 
In particular the extracted value of the polar angle $\theta$ is found
to show rather large variation with $S_{low}$ for the case of number 
counts as well as number counts weighted by degree of polarization. 
This variation can be attributed to the presence of $\theta$
dependent bias in data.  

The variation in $\theta$ in the case of number counts weighted by 
polarized flux density, however, is found to be 
relatively small. This motivates
us to study this observable in greater detail. This observable is also
physically relevant since it represents the total polarized flux in each
pixel. We study the dipole anisotropy in this observable for several
different different cuts which eliminates data with large error in 
polarized flux density. We also determine the change in signal with
increase in the lower limit on the polarized flux density. We find that
the dipole amplitude in this case is relatively stable with these changes.
The results for different cuts agree within errors.   
The extracted azimuthal angle of the dipole axis is also found to be stable.
The maximum change is seen in the polar angle, $\theta$. However in
this case also the result stabilizes as we go to larger values of $S_{low}$.

The observed anisotropy cannot be attributed to the 
local supercluster. In our study we remove  
sources lying in our local 
neighbourhood. These sources are identified by using the catalogues 
provided in 
\citep{Saunders:2000,Vaucouleurs:1991,Corwin:1994}. Alternatively we have also studied
the signal after eliminating the supergalactic plane. The results do not
depend very strongly on these cuts.

Finally, we calculate the velocity for the case of number
counts weighted by polarized flux density.
Here we assume that the
polarized flux follows a power law dependence on frequency ($P \propto\nu^{-\alpha_P}$), similar to intensity. 
The extracted speed turns out to be even larger in comparison to that
extracted without imposing polarization cuts
 \citep{Singal:2011,Kothari:2013}. In this analysis we assume that 
$\alpha_P\approx \alpha$, the flux spectral index. This value may be
subject to change once better polarization spectral data becomes available.

We conclude that the anisotropy in polarization
observables is much stronger in comparison to flux observables.   
This is consistent with the earlier claim of significant dipole anisotropy
in the polarization offset angle \citep{Jain:1998r}. 

There now exist many independent observations which indicate a preferred 
direction pointing roughly towards Virgo. It is unlikely that all of
them can be explained by some systematic effect. For example, 
the NVSS data is more likely to pick a preferred axis pointing towards
the poles due to systematic effect arising due to sources with low
flux \citep{Blake:2002}. The direction observed, however, is nearly 
perpendicular to that. The dependence of direction with the cut on flux
density might be explained by this systematic effect. This dependence
is most prominent in number counts but is relatively small in the
flux weighted number counts \citep{Kothari:2013}  
as well as polarized flux weighted 
number counts. Furthermore it seems very unlikely that 
systematic effects would pick the same direction in so many different
observations, i.e. radio polarizations orientations \citep{Jain:1998r},
optical polarizations \citep{Hutsemekers:1998}, CMBR quadrupole and octopole
 \citep{Tegmark:2004}, radio number counts \citep{Blake:2002,Singal:2011}
and radio polarized flux (present work). 
In all likelihood this alignment of axes \citep{Ralston:2004,Schwarz:2004} is caused
by a physical effect. There exist many models which might explain the
observed large scale anisotropy. An interesting possibility, within the
framework of the Big Bang model, is that the modes generated during the
pre-inflationary anisotropic expansion might enter the horizon at late times
and cause to observe large scale anisotropy \citep{Aluri:2012,Rath:2013}. The early phase of expansion
is expected to be anisotropic, which evolves into an isotropic Universe 
during inflation \citep{Wald:1983}. The large scale anisotropy may 
also affect the intergalactic magnetic field which may influence the 
source magnetic field and hence its 
polarized radiation.  
A detailed investigation of the physical mechanism is postponed 
to future work. 

\section*{Acknowledgements}
We have used CERN ROOT 5.27 for generating our plots. Some of the results in this paper have 
been derived using the HEALPix \citep{Gorski:2005} package. 
Prabhakar Tiwari sincerely acknowledge CSIR, New Delhi for award of fellowship during the work. We thank William Cotton for useful comments on the paper. 
\bibliographystyle{mn2e}
\bibliography{radioPol}
\end{document}